\documentclass{article}
\usepackage{arxiv_preprint,times}
\usepackage[T1]{fontenc}
\usepackage[utf8]{inputenc}
\input{glyphtounicode}
\usepackage{microtype}
\usepackage{amsmath,amssymb}
\usepackage{graphicx}
\usepackage{booktabs}
\usepackage{tabularx}
\usepackage{longtable}
\usepackage{array}
\usepackage{enumitem}
\usepackage{titlesec}
\usepackage{xcolor}
\usepackage{seqsplit}
\usepackage{needspace}
\usepackage{placeins}
\usepackage{tikz}
\usetikzlibrary{arrows.meta,positioning,calc,fit}
\usepackage[colorlinks=true,allcolors=blue]{hyperref}
\usepackage{xurl}
\hypersetup{
  pdftitle={What Does an Evaluation License? A Commit-Bound Census of Claim Replay in Inspect Evals},
  pdfauthor={Xi Qin; Jizhou Tong},
  bookmarksopen=true,
  bookmarksnumbered=true
}
\titlespacing*{\section}{0pt}{1.25ex plus .3ex minus .2ex}{.62ex}
\titlespacing*{\subsection}{0pt}{.95ex plus .2ex minus .15ex}{.42ex}
\titlespacing*{\paragraph}{0pt}{.7ex plus .2ex minus .1ex}{.5em}
\newcommand{\pp}{\,pp}
\newcommand{\STOP}{\textsc{stop}}
\newcommand{\ID}{\textsc{id}}
\newcommand{\NID}{\textsc{non-id}}
\newcommand{\F}{\mathcal F}
\newcommand{\Fpri}{\F_{\mathrm{primary}}}
\newcommand{\Frev}{\F_{\mathrm{review}}}
\newcommand{\Fdisp}{\F_{\mathrm{disputed}}}
\newcommand{\I}{\mathcal I}
\newcommand{\relset}{\mathcal R}

\newcolumntype{Y}{>{\raggedright\arraybackslash}X}

\title{What Does an Evaluation License?\\A Commit-Bound Census of Claim Replay\\in Inspect Evals}
\author{
Xi Qin\\
Wuhan University
\and
Jizhou Tong\\
Wuhan University
}
\date{}

\begin{document}
\maketitle
\suppressfloats[t]

\begin{abstract}
\textbf{Benchmarks can run without determining what their results license.}  We freeze a large evaluation collection and attempt to replay its historical claims.  Most units stop because the evidence required for replay is not bound.  Where replay is possible, different claims remain stable at different resolutions.  We make this otherwise implicit inference step explicit and executable.  Concretely, we take a commit-bound census of 124 eligible units in Inspect Evals, a large community collection for the Inspect AI framework: 110 stop and 14 permit deterministic analysis.  Across the completed analyses, scores and full rankings can vary while winners and many pairwise relations remain fixed.  Standard identified sets over frozen evidence \(D\), an admitted evaluator family \(\F\), and a stated claim \(q\) return explicit stopping reasons, disagreement witnesses, and stable comparisons.
\end{abstract}

\begin{center}
\fcolorbox{blue!55!black}{blue!4}{%
\begin{minipage}{.94\linewidth}
\small
\textbf{One result, two documented readings.}
On the same AgentDojo attack outcomes, the published-micro reading ranks Gemini ahead of GPT-4o-mini (20.827 vs.\ 27.186; lower is better), while the utility-conditional reading reverses them (32.000 vs.\ 19.745).  Claude remains first under both.  The score and lower ordering change; the winner does not.  \textbf{One census, one hard boundary.} In the complete 124-unit frame, 110 units lack evidence required to reconstruct a historical claim and stop with an explicit reason.  The paper asks one simple question: for a stated claim, what remains true across the evaluator meanings admitted to the analysis?
\end{minipage}%
}
\end{center}

\fontsize{10}{11.6}\selectfont
\setlength{\parskip}{1.2pt plus .3pt minus .2pt}
\setlength{\textfloatsep}{7pt plus 2pt minus 1pt}
\setlength{\floatsep}{6pt plus 2pt minus 1pt}
\setlength{\intextsep}{6pt plus 2pt minus 1pt}
\setlength{\abovedisplayskip}{5pt plus 1pt minus 1pt}
\setlength{\belowdisplayskip}{5pt plus 1pt minus 1pt}

\section{Introduction: running code is not replaying a claim}

Modern evaluation infrastructure increasingly provides task registries, scorers, structured logs, containers, and leaderboards.  Yet the scientific inference attached to those executions often remains ambient: it depends on repository layout, maintainer knowledge, \texttt{README} conventions, implicit missingness rules, and manual evidence assembly.  Systems engineering addresses analogous risks through hermetic builds, capability-bounded interfaces, supply-chain attestations, and fail-closed verification.  Claim replay brings the same discipline to evaluation conclusions: evidence outside the bound interface may exist, but it does not silently license inference.  \textbf{Operationally, this requires a \emph{claim-replay contract} binding the historical observations, the admissible semantic alternatives, and the claim being queried.}

We audit a commit-bound finite frame within Inspect Evals, the community-contributed evaluation collection and register for the Inspect AI framework.  Inspect is a frontier-AI evaluation framework developed by the UK AI Security Institute and Meridian Labs, with over 200 pre-built evaluations spanning coding, agentic tasks, reasoning, knowledge, behavior, and multimodal understanding \citep{InspectFramework2026}.  Inspect Evals was created by UK AISI, Arcadia Impact, and the Vector Institute; since 8 May 2026, new contributions have been hosted in upstream repositories and listed through a distributed register \citep{InspectEvalsRepository2026,InspectEvalsRegister2026}.

At the frozen snapshot, mechanical eligibility yields 124 evaluation units.  This is the complete eligible frame at that commit, not today's full distributed Inspect ecosystem.  The pinned repository commit is how we freeze the frame for reproducibility; Inspect Evals itself is the scientific population being positioned.

AgentDojo then shows the claim problem before any formalism.  Two documented readings of the same attack outcomes reverse the Gemini/GPT-4o-mini ordering, yet both retain Claude as the winner.  This is not a contradiction: ``Who wins?'' and ``What is the full ranking?'' are different claims.  The audit first asks whether a historical claim can be reconstructed from bound evidence.  When it can, we ask which answers agree across the primary analysis and a wider review family.

The same issue appears under tighter control in Appendix~\ref{app:legacy-worked-examples}: with observations and predictions fixed, changing only one terminal rule moves an estimated effect from about \(-40\) percentage points to zero.  These examples create the need for a precise object.  Section~\ref{sec:formal-layer} introduces it only after the phenomenon is visible.

\begin{figure}[!b]
\centering
\resizebox{\linewidth}{!}{%
\begin{tikzpicture}[
  font=\scriptsize,
  >=Latex,
  stage/.style={draw,rounded corners=2pt,align=center,text width=3.25cm,minimum height=1.45cm,inner sep=4pt,line width=.55pt},
  result/.style={draw,rounded corners=2pt,align=center,text width=3.25cm,minimum height=1.22cm,inner sep=4pt,line width=.55pt},
  arrow/.style={->,line width=.7pt}
]
\node[stage,fill=blue!7] (frame) at (0,1.9) {\textbf{Inspect Evals finite frame}\\124 eligible units\\frozen at one commit};
\node[stage,fill=orange!10] (gate) at (4.25,1.9) {\textbf{Is the evidence bound?}\\historical observations, traces, state, support};
\node[stage,fill=green!9] (family) at (8.50,1.9) {\textbf{Which meanings count?}\\same-claim/support screen\\\emph{primary / review families}};
\node[stage,fill=purple!8] (enum) at (12.75,1.9) {\textbf{What remains true?}\\values, winners, orders,\\stable pairs};
\draw[arrow] (frame) -- (gate);
\draw[arrow] (gate) -- (family);
\draw[arrow] (family) -- (enum);
\node[result,fill=orange!6] (stop) at (4.25,0) {\textbf{Cannot replay claim}\\explicit stopping reason\\missing item and minimum unblock};
\node[result,fill=green!6] (mult) at (8.50,0) {\textbf{Several admitted readings}\\primary conclusions\\and wider review sensitivity};
\node[result,fill=purple!6] (sep) at (12.75,0) {\textbf{Claim-specific answers}\\a value may change while\\a winner or pair stays fixed};
\draw[arrow] (gate) -- (stop);
\draw[arrow] (family) -- (mult);
\draw[arrow] (enum) -- (sep);
\node[align=center,font=\tiny] at (8.5,-1.02) {Humans decide which meanings count; calculation and verification after that decision are deterministic and fail closed.};
\end{tikzpicture}
}
\caption{\textbf{Three questions, three outputs.} Missing evidence, multiple evaluator meanings, and disagreement across claim levels are not one generic robustness outcome.}
\label{fig:missing-layer}
\end{figure}

\paragraph{Contributions.}
We contribute three linked pieces:
\begin{enumerate}[leftmargin=*,label=(\roman*),itemsep=.15ex,topsep=.25ex]
  \item a claim-replay formulation that makes ``what remains true?'' executable using standard identified sets over frozen evidence and admitted evaluator meanings;
  \item a complete 124-unit census showing which historical claims can be replayed and the explicit reason when they cannot; and
  \item a practical audit and release interface that reports values, directions, winners, full orders, and stable pairwise relations separately.
\end{enumerate}
The paper follows Figure~\ref{fig:missing-layer} from left to right: Section~2 moves from the AgentDojo reversal to the formal object, Section~3 asks whether the historical evidence is sufficient, Section~4 reports what survives across evaluator meanings, and Section~5 states the release interface.

\paragraph{Public artifact entry point.}
{\raggedright The arXiv source includes \texttt{anc/claim\_replay\_public\_reproduction.zip}, one portable package covering the census, complete audits, and three deterministic micro-experiments.  Start at its \texttt{README.md}; the top-level verifier checks package closure, release-time and local-path hygiene, the claim-replay verifier, deterministic regeneration, and the independent 844-check path.\par}

\section{From one ranking reversal to an identified set}
\label{sec:formal-layer}

\subsection{The phenomenon before the notation}

AgentDojo holds the attack outcomes fixed but permits several documented ways to decide eligibility, denominators, and aggregation.  The published-micro and all-run-micro readings form the primary analysis.  Three additional readings are plausible enough to inspect but disputed, so they enter only the wider review family.

\begin{table}[ht]
\centering
\caption{AgentDojo targeted attack success rate (lower is better) over the same outcomes.  The primary analysis uses the first two readings; the review family also asks what happens under the three disputed readings.}
\label{tab:agentdojo-worked}
\footnotesize
\setlength{\tabcolsep}{3.2pt}
\begin{tabular}{llrrr}
\toprule Reading & Family & Claude 3.5 & Gemini 2.0 & GPT-4o-mini \\
\midrule
Published micro & Primary & 1.113 & 20.827 & 27.186 \\
All-run micro & Primary & 1.113 & 14.120 & 27.186 \\
Common-suite macro & Review only & 1.473 & 26.731 & 30.665 \\
Utility-conditional & Review only & 1.316 & 32.000 & 19.745 \\
Worst suite & Review only & 3.810 & 66.667 & 57.143 \\
\addlinespace
\multicolumn{5}{l}{\textit{Primary (2A): one winner; one full order; 3/3 pairs agree}} \\
\multicolumn{5}{l}{\textit{Review (2A+3D): one winner; two full orders; 2/3 pairs agree}} \\
\bottomrule
\end{tabular}
\end{table}

In the primary analysis, Claude wins, the full order is unique, and all three pairwise relations agree.  In the wider review family, Claude still wins, but two readings place GPT-4o-mini above Gemini.  The full ranking is no longer unique, although Claude's two relations remain fixed.  Published micro versus utility-conditional is enough to display the disagreement.  A single flip/no-flip label would hide both the stable winner and the fact that the reversal enters only under the wider family.

This example creates a precise question without requiring new vocabulary: for a fixed claim---a score, direction, winner, or full order---do all evaluator meanings in the chosen family give the same answer?  The percentages remain attached to each reading because their eligibility and aggregation differ; we do not pool them into one exact-score estimand.  Full provenance and support decisions are in Appendix~\ref{app:r8-agentdojo-v0-1-35}.

\subsection{The minimal formal object}

Let \(D\) denote the frozen evidence needed for the claim: system outputs together with the relevant sample, judge, state, support, and precision information.  An evaluator specification \(f\) maps that same evidence to a typed value \(V(f;D)\).  Reviewers classify each candidate as admitted for the primary analysis (A), disputed but relevant for review (D), or excluded (X).  This yields
\begin{equation}
  \Fpri=\{f:a(f)=\mathrm A\},\qquad
  \Frev=\Fpri\cup\Fdisp,\quad \Fdisp=\{f:a(f)=\mathrm D\}.
  \label{eq:admission-families}
\end{equation}
The primary family therefore states the paper's main interpretation; the review family shows sensitivity to documented alternatives that did not clear primary admission.

For claim \(q\), collect every answer allowed by the chosen family:
\begin{equation}
  \I_q(D,\F)=\{q(V(f;D)):f\in\F\}.
  \label{eq:identified-set}
\end{equation}
We call \(\I_q(D,\F)\) the \emph{claim-relative identified set}.  The claim is identified when this set has one member.  The query \(q\) may ask for an exact value, a direction or threshold crossing, a winner/top set, or a complete weak order.  Here \emph{exact} means exact for the declared frozen evidence \(D\); a published rounded table does not reveal latent unrounded values.  Exact-value sets are formed only for commensurate scalar endpoints.  Ordinal analyses may display values per specification without pooling their magnitudes.

For systems \(a,b\), we also collect their relation under every admitted meaning:
\begin{equation}
  \relset_{ab}(D,\F)
  =\{\operatorname{rel}(V_a(f;D),V_b(f;D)):f\in\F\}.
  \label{eq:pairwise-image}
\end{equation}
The pair is stable when this set has one member.  If a claim is not identified, two specifications with different answers provide a concrete witness; the singleton pairwise relations state exactly which comparisons still hold.

\subsection{Two choices that must stay separate}

First, reviewers choose which evaluator meanings enter \(\F\).  A candidate must address the same claim on comparable support and must state its endpoint, missingness rule, aggregation, unit, orientation, and tie policy.  Source exposure alone does not guarantee admission, and a broken evaluator is recorded as a reliability problem rather than treated as another meaning.  Appendices~\ref{app:technical-formalism} and~\ref{app:executable-contract} give the full review and evidence rules.

Second, the claim \(q\) sets the level of resolution.  Adding evaluator meanings can only enlarge the answer set:
\begin{equation}
  \F\subseteq\F'
  \quad\Longrightarrow\quad
  \I_q(D,\F)\subseteq\I_q(D,\F').
  \label{eq:domain-monotonicity}
\end{equation}
So a conclusion fixed in the primary family may vary in the wider review family.  Conversely, if \(q_2=h\circ q_1\) asks a coarser question, agreement on the finer claim implies agreement on the coarser one, but not the reverse.  A full order can vary while its winner and many pairwise relations remain fixed.  Evaluator variation, sampling uncertainty, printed precision, and reversible unit changes remain separate objects (Appendix~\ref{app:technical-formalism}).

\section{Can the historical claim be replayed?}

\subsection{The complete frame and explicit stops}

For reproducibility, the census is bound to one frozen Inspect Evals repository commit.  At public commit \texttt{32b79a2}, a predeclared mechanical eligibility predicate yields 124 units from 129 strict-path candidates in the frozen tree.  Five rule-based exclusions are fully enumerated in Appendix~\ref{app:frame-predicate}; none depends on an evaluation outcome.  This commit defines the audited finite frame, not the current distributed register.

\textbf{Every eligible unit receives a final record.  Fourteen contain enough bound evidence for deterministic claim analysis; 110 stop earlier.}  We write \STOP{} for this explicit stopping reason, not for a zero score or a failed benchmark.  Of the 110 stops, 103 belong to the outcome-blind census-review path and seven to the frozen random draw.  In the former, the leading reasons are evidence-binding failure (47), missing comparative observations (35), and missing judge traces (9); the remaining twelve concern endpoint/target definition, missingness, claim selection, or terminal state.  Each record names the first unavailable item and the minimum artifact that would permit claim replay.  Appendix~\ref{app:frame-accounting} gives the full taxonomy and accounting.

\subsection{What the three evidence streams can say}

The census and the worked analyses answer different questions.  The census counts which units in the complete pinned frame can reach claim analysis.  A separately frozen random draw probes completion under one acquisition protocol.  Purposive and outcome-exposed cases explain mechanisms and broaden the set of examples.  Table~\ref{tab:accounting-licenses} states these limits directly.

\begin{table}[t]
\centering
\caption{Three evidence streams and the statements they license.}
\label{tab:accounting-licenses}
\small
\setlength{\tabcolsep}{4pt}
\begin{tabularx}{\linewidth}{@{}p{0.20\linewidth}p{0.30\linewidth}Y@{}}
\toprule
Evidence stream & Design & What it licenses \\
\midrule
124-unit census & Complete mechanically eligible frame at one commit & Within-frame accounting: 14 complete, 110 explicit \STOP{} records \\
Frozen random ten & Two per stratum; case selection frozen before terminal audit outcomes; outcome exposure recorded separately & Completion under that frozen balanced-strata draw: 3/10 \\
Mechanism portfolio & Three purposive deep audits plus eight labeled outcome-exposed discoveries & Examples of how claims agree or disagree, not a frequency \\
\bottomrule
\end{tabularx}
\end{table}

The complete census supports within-frame accounting; the random ten supports its own frozen balanced-strata statement; and the mechanism portfolio supports case-level explanation.  \textbf{These streams are never pooled, and none is an ecosystem-prevalence estimate.}  Appendix~\ref{app:frame-accounting} gives the estimator, outcome-exposure states, and full accounting rules.

\subsection{A fresh run answers a new question}

A fresh model, judge, agent, or sandbox run changes \(D\).  It may support a valuable prospective study, but it cannot reconstruct the historical event.  The 110 stops therefore remain in the census rather than being replaced by current outputs.  Judge, sandbox, dependency, and terminal-state examples are detailed in Appendix~\ref{app:preservation-layers}.

\subsection{Why task code and logs are not enough}

Inspect already exposes structured task, scorer, execution, registry, and \texttt{EvalLog} machinery for forward evaluation and rich sample-level history.  Historical claim replay adds a narrower requirement: a published claim must travel with the least artifact sufficient to identify its historical instance, admitted evaluator meanings, and question:
\[
\text{task/scorer/log}\;\not\Rightarrow\;(D,\F,q)_{\mathrm{executable}}
\;\Longrightarrow\;\I_q(D,\F)
\]
\textbf{The first arrow is not automatic: the release must bind the historical evidence, admitted evaluator family, and claim.  Once those are fixed, the second arrow is deterministic.}  Table~\ref{tab:native-replay-contract} states the distinction; Appendix~\ref{app:native-log-binding} provides the source-level audit.

\begin{table}[t]
\centering
\caption{What native evaluation objects provide, and what a historical claim still needs.}
\label{tab:native-replay-contract}
\small
\begin{tabularx}{\linewidth}{@{}p{0.19\linewidth}p{0.34\linewidth}Y@{}}
\toprule Native object & Already supports & What the historical claim still needs \\
\midrule
task/scorer & Forward task and scoring definition & Historical instance and replay target \\
\texttt{EvalLog}, when retained & Inputs, outputs, targets, scores, metadata, reductions & Support decisions and admitted evaluator family \\
report & Metrics and reproducibility metadata & The claim being asked and the answers that agree or disagree \\
\bottomrule
\end{tabularx}
\end{table}

Across the 124 records, task/scorer availability repeatedly precedes missing historical observations, judge decisions, state, support joins, or an evaluator registry.  Thus the audited artifacts often support forward execution without carrying the last links needed to reconstruct a published conclusion.  A compact sufficient statistic can close a coarse claim; a stateful endpoint may require restorable state.  The preservation matrix is in Appendix~\ref{app:preservation-layers}.

\section{What remains true when replay is possible}

\subsection{Results across claims and evaluator families}

Once the historical evidence \(D\) is available, we compute the answer under every evaluator meaning admitted to the chosen family.  The answers need not all vary together.  A score can change while the winner stays fixed; a full order can change while most pairwise comparisons remain fixed.  The identified set records this separately for each claim.

This subsection keeps three evidence streams separate: three completed random draws describe the frozen draw, three purposive audits expose mechanisms under named families, and eight outcome-exposed cases map breadth.  One purposive case, BEIR, also receives a separate 25-view coverage diagnostic.  None of these results is pooled.  Table~\ref{tab:mechanism-matrix} gives the main-text map; case-level values and the eight exploratory rows remain in Appendices~\ref{app:complete-audit-records} and~\ref{app:exploratory-eight}.

\begin{table}[t]
\centering
\caption{What stays fixed across evaluator meanings, separated by evidence stream.  P/E denote the primary/review families \(\Fpri/\Frev\).  Detailed random and exploratory rows are in the appendix.}
\label{tab:mechanism-matrix}
\footnotesize
\setlength{\tabcolsep}{2.8pt}
\begin{tabularx}{\linewidth}{@{}p{0.18\linewidth}p{0.30\linewidth}p{0.19\linewidth}Y@{}}
\toprule
Stream / case & Evaluator choices & P/E winner or direction & P/E order / stable pairs \\
\midrule
Random (3 complete) & credit, aggregation, claim target & all: \NID/\NID & all: \NID/\NID; 0/1, 0/1, 0/3 in both \\
\addlinespace
Purposive: AgentDojo & eligibility, denominator, aggregation & winner \ID/\ID & P: \ID, 3/3; E: \NID, 2/3 \\
Purposive: AutoML & missingness, task/rank aggregation & winner \ID/\ID & P: \ID, 10/10; E: \NID, 9/10 \\
Purposive: BEIR (5 views) & metric, cutoff, relevance grading & winner \ID/\ID & \NID/\NID; 2/3 in both \\
\addlinespace
Outcome-exposed exploratory (8) & source/evaluator readings and audit proposals & \multicolumn{2}{l}{Case-specific landscape; no pooled P/E claim} \\
\bottomrule
\end{tabularx}
\end{table}

\paragraph{Completed random draws.}
All three completed random draws yield more than one winner and more than one full order in both families.  Their mechanisms differ---credit definition, aggregation, and claim target---and their full answer sets are reported in Appendix~\ref{app:complete-audit-records}.  Randomization licenses the completion-rate statement for the frozen draw; it does not license a prevalence estimate for evaluator disagreement, because subsequent evaluator adjudication was not outcome-blind.

\paragraph{Purposive audits under the named families.}
Expanding from the primary to the review family makes the complete orders for AgentDojo and AutoML vary, yet no previously identified winner is lost.  Across the three purposive audits, 13 of 15 previously identified pairwise relations remain identified.  Adding evaluator meanings therefore need not erase every conclusion at once.  Broader candidate generation yielded no additional executable specifications in these six audits.

\paragraph{Separate BEIR coverage diagnostic.}
The 25-view BEIR analysis is not part of the named-family result above.  It is an \emph{exploratory coverage sensitivity}: the winner varies and stable pairs fall from 2/3 to 0/3.  This does not license the enlarged grid as a completed or admitted family.

\paragraph{Outcome-exposed exploratory cases.}
The eight outcome-exposed cases range from representation-only changes to exact-value sensitivity, lower-pair reversals, and a support-limited candidate winner change.  Their purpose is breadth and boundary testing, not a pooled headline.  Appendix~\ref{app:exploratory-eight} preserves every case, admission decision, and support limitation.

The shared lesson is positive as well as negative.  \textbf{A fine claim can vary while a coarser direction or winner and many pairwise comparisons remain fixed.}  This result is always relative to the named \((D,\F_{\mathrm{scope}},q)\): a flip-only report discards useful structure, while extending the conclusion beyond the named family overstates it.

\begin{table}[t]
\centering
\caption{What an available-case sensitivity summary can lose, and what claim replay adds.}
\label{tab:added-information}
\footnotesize
\setlength{\tabcolsep}{3pt}
\begin{tabularx}{\linewidth}{@{}p{0.20\linewidth}p{0.34\linewidth}Y@{}}
\toprule Question & Available-case sensitivity summary & Claim-replay result \\
\midrule
Can the claim be replayed? & Conditions on cases with usable observations or leaves missing cases outside the numerical analysis & Retains all 124 records; 110 receive an explicit \STOP{} with the first missing item and minimum unblock \\
Which choices define the claim? & May pool all executable variants into one choice set & Separates the primary family from the wider review family; AgentDojo and AutoML have one primary order but multiple review-family orders \\
What survives variation? & Often reports a scalar range or any flip & Reports claim-specific answer sets, witnesses, and stable relations; the BEIR sensitivity shows that 2/3 stable pairs over five views can become 0/3 over an expanded evaluator grid \\
\bottomrule
\end{tabularx}
\end{table}

Thus a naive complete-case multiverse would make two distinct errors here: it would omit cases with missing evidence, and it could collapse primary and disputed meanings into one enlarged family.  A flip-only summary would then discard the fixed winner and stable comparisons that remain decision-relevant.

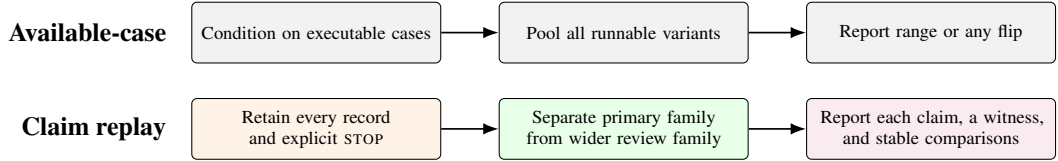
\begin{figure}[t]
\centering
\resizebox{\linewidth}{!}{%
\begin{tikzpicture}[
  font=\scriptsize,
  >=Latex,
  box/.style={draw,rounded corners=2pt,align=center,minimum height=.82cm,text width=3.15cm,inner sep=3pt},
  arrow/.style={->,line width=.65pt}
]
\node[box,fill=gray!10] (avail1) at (0,1.15) {Condition on executable cases};
\node[box,fill=gray!10] (avail2) at (4.15,1.15) {Pool all runnable variants};
\node[box,fill=gray!10] (avail3) at (8.30,1.15) {Report range or any flip};
\draw[arrow] (avail1) -- (avail2);
\draw[arrow] (avail2) -- (avail3);
\node[anchor=east,font=\bfseries] at (-1.9,1.15) {Available-case};

\node[box,fill=orange!10] (typed1) at (0,-.15) {Retain every record\\and explicit \STOP};
\node[box,fill=green!9] (typed2) at (4.15,-.15) {Separate primary family\\from wider review family};
\node[box,fill=purple!8] (typed3) at (8.30,-.15) {Report each claim, a witness,\\and stable comparisons};
\draw[arrow] (typed1) -- (typed2);
\draw[arrow] (typed2) -- (typed3);
\node[anchor=east,font=\bfseries] at (-1.9,-.15) {Claim replay};
\end{tikzpicture}%
}
\caption{\textbf{What the added interface changes.} It preserves cases that cannot be replayed, separates primary from disputed meanings, and states what remains fixed alongside what changes.}
\label{fig:available-case-contrast}
\end{figure}

\subsection{Same predictions, different evaluator meaning}

\paragraph{Changing one rule changes the conclusion.}
This orthogonal 217-row diagnostic dataset is outside the Inspect census denominator.  Across the controlled comparison, the rows, support, and seven prediction rules are fixed; only the terminal rule changes.

\begin{figure}[t]
\centering
\resizebox{.98\linewidth}{!}{%
\begin{tikzpicture}[font=\scriptsize,>=Latex]
  \fill[blue!3,rounded corners=3pt] (-.30,-.65) rectangle (6.05,6.05);
  \fill[purple!3,rounded corners=3pt] (6.45,-.65) rectangle (13.00,6.05);

  \node[anchor=west,font=\bfseries\large] at (0,5.62) {A\quad Effect collapse};
  \node[anchor=west,text=gray!65!black] at (0,5.20) {change only the terminal rule};
  \node[draw=gray!45,fill=white,rounded corners=2pt,align=center,
        minimum width=5.15cm,minimum height=.62cm] (fixed) at (3.00,4.65)
        {Fixed: 217 rows \quad support \quad seven prediction rules};

  \draw[->,gray!70] (.45,.38) -- (.45,3.78);
  \foreach \y/\lab in {.65/{\(-40\)},2.05/{\(-20\)},3.45/{\(0\)}} {
    \draw[gray!55] (.36,\y) -- (.54,\y);
    \node[anchor=east] at (.30,\y) {\lab\pp};
  }
  \draw[gray!30] (.55,3.45) -- (5.72,3.45);
  \draw[very thick,red!70!black] (1.35,.52) -- (3.20,3.45) -- (5.05,3.45);
  \fill[red!75!black] (1.35,.52) circle (3.3pt);
  \fill[green!55!black] (3.20,3.45) circle (3.3pt);
  \fill[green!55!black] (5.05,3.45) circle (3.3pt);
  \node[anchor=west,font=\bfseries,red!70!black,fill=blue!3,inner sep=1.5pt]
        at (2.05,1.10) {\(\approx-40\)\pp};
  \node[font=\bfseries,green!45!black] at (3.20,3.78) {\(0\)\pp};
  \node[font=\bfseries,green!45!black] at (5.05,3.78) {\(0\)\pp};
  \node[align=center] at (1.35,-.12) {Original\\asymmetric};
  \node[align=center] at (3.20,-.12) {Retain\\both};
  \node[align=center] at (5.05,-.12) {Exclude\\both};

  \node[anchor=west,font=\bfseries\large] at (6.78,5.62) {B\quad Which claim is asked?};
  \node[anchor=west,text=gray!65!black] at (6.78,5.20) {same exclude-both specification cells};
  \node[draw=blue!60!black,fill=blue!5,rounded corners=3pt,align=center,
        minimum width=4.10cm,minimum height=1.05cm] (effect) at (9.73,4.32)
        {\textbf{Adoption effect}\quad \textsc{id}: \(0\)\pp};
  \node[draw=red!65!black,fill=red!5,rounded corners=3pt,align=center,
        minimum width=2.42cm,minimum height=1.55cm] (order) at (8.15,2.50)
        {\textbf{Complete ranking}\\[2pt]{\large\textsc{non-id}}\\5 weak orders};
  \node[draw=green!55!black,fill=green!6,rounded corners=3pt,align=center,
        minimum width=2.42cm,minimum height=1.55cm] (winner) at (11.35,2.50)
        {\textbf{Winner}\\[2pt]{\large\textsc{id}}\\Mini in every cell};
  \draw[->,line width=.8pt] (order.east) -- (winner.west);
  \node[align=center,text width=5.7cm] at (9.73,.65)
        {The same cells answer three different questions; one answer can be fixed while another varies.};
\end{tikzpicture}%
}
\caption{\textbf{One rule changes the conclusion while predictions stay fixed.} Left: replacing the original one-sided terminal rule with either symmetric rule moves the adoption effect from approximately \(-40\)\pp{} to zero. Right: the exclude-both family still permits five weak orders, yet Mini is the unique winner.}
\label{fig:controlled-semantic-main}
\end{figure}
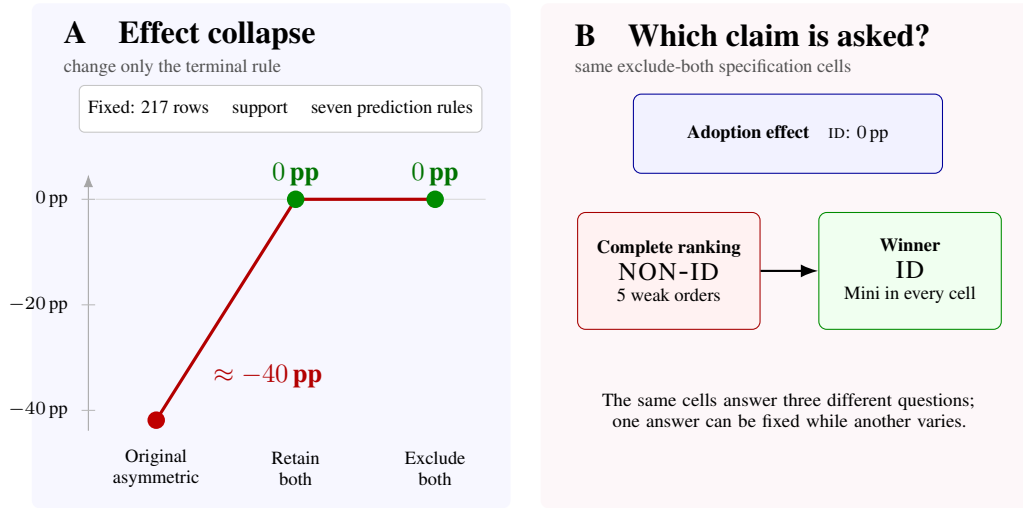
\FloatBarrier

Exact original effects span \(-43.03\) to \(-40.76\)\pp{} across the three disposition policies.  Their collapse from approximately \(-40\)\pp{} to zero is not ordinary within-specification sampling uncertainty: the rows and predictions have not changed, but the conclusion has.  The omitted terminal rule therefore has a scientific consequence rather than merely changing how uncertainty around one fixed specification is summarized.

The right panel makes a second point visible.  Under exclude-both, the adoption effect is exactly zero, the complete ranking spans five weak orders, and Mini remains the unique winner in every cell.  These statements do not conflict: effect, full order, and winner are different questions about the same observations.  A claim-replay result must therefore say which question it answers instead of reporting one unqualified robust/not-robust label.

\paragraph{What this comparison isolates.}
The original asymmetric rule treats a terminal prediction differently on the two contrast sides; retain-both and exclude-both restore comparability in opposite ways.  Both symmetric readings return zero, localizing the approximately \(-40\)\pp{} conclusion to the one-sided rule while the recorded predictions remain fixed.  Appendix~\ref{app:legacy-worked-examples} gives the full control design and its causal and transport limits.

\paragraph{Why the winner still matters.}
Zero adoption effect does not imply that every ranking claim has one answer.  The exclude-both cells still support five complete weak orders because lower-ranked systems exchange positions across admitted specifications.  Yet all those orders place Mini first.  Reporting only ``ranking varies'' would therefore conceal a decision-relevant winner.  The appropriate output gives a separate answer for effect, direction, winner, complete order, and stable pairs.  This is the same principle used in the Inspect audits, demonstrated here while the observations remain fixed.

Figure~\ref{fig:controlled-semantic-main} is the main-text view.  Appendix~\ref{app:legacy-worked-examples} retains the complete intervals, four-corner decomposition, empty-family witness, cross-corpus heterogeneity, full rankings, and study boundaries.

\section{A practical claim-replay interface}

The interface has one deliberate human boundary.  Human auditors define the claim, endpoint type, comparable support, candidate origin, and A/D/X admission; prospective audits freeze those choices before outcomes.  \textbf{Human judgment ends there.}  Automation is not asked to discover the uniquely correct evaluator meaning or certify that the family is complete.

\begin{center}
\footnotesize
\setlength{\tabcolsep}{3pt}
\begin{tabularx}{.98\linewidth}{@{}p{.20\linewidth}Y@{}}
\toprule \textbf{Field group} & \textbf{What the release binds} \\
\midrule
Claim/evidence & claim and query IDs; endpoint type; systems, observations, and source/state bindings \\
Candidate ledger & definition and origin; reliability; A/D/X verdict, rationale, and same-target/support decision \\
Support/aggregation & eligibility, denominator, missingness/DNF, aggregation tree, weights, units, orientation, and ties \\
Exposure/freeze & selection and exposure state; evidence horizon; pre-outcome ledger digest; chained amendments \\
Output/\STOP & primary/review answers, a disagreement witness, and stable pairs; or an explicit stop reason and minimum unblock \\
\bottomrule
\end{tabularx}
\end{center}

After admission is frozen, execution is deterministic.  The analyzer binds \(D\), places A in \(\Fpri\) and A+D in \(\Frev\), evaluates each specification, and computes values, top sets, weak orders, pairwise relations, and minimum disagreement witnesses in both families.  Missing observations, support decisions, orientation, or admission produce an explicit \STOP{} with an actionable unblock rather than imputation.  In a controlled evidence-ablation test over five otherwise complete packets, all five intact packets remained complete, while all 20 targeted evidence deletions stopped exactly at the declared G1--G4 boundary (Appendix~\ref{app:microexperiments}).

An independent implementation recomputes the arithmetic and answer sets from the bound source tables rather than reading stored discovery outputs.  It reproduces fractions, orders, stable pairs, and support boundaries; the provenance fields remain public for human audit.  Appendix~\ref{app:independent-recomputation} gives the arithmetic chains, and Appendix~\ref{app:executable-contract} states the full validation boundary.

\subsection{What a reader sees}

The audit does not compress everything into one robustness label.  It reports separate answers for the value, direction or threshold, winner/top set, complete order, and pairwise comparisons.  Each claim with more than one answer carries a two-specification witness; each stable comparison states what survived.  The observed disagreement is also not generally attributable to one challenger: it persists under 71 of 76 leave-one-out removals, and admitted--admitted pairs alone furnish 14 distinct disagreement witnesses (Appendix~\ref{app:microexperiments}).

Three reporting rules keep that output interpretable:
\begin{enumerate}[leftmargin=*,itemsep=.15ex,topsep=.25ex]
  \item name \((D,\F_{\mathrm{scope}},q)\) and report primary and review-family conclusions separately;
  \item give every \STOP{} its missing item and minimum unblock, never a numeric surrogate; and
  \item retain case selection, outcome exposure, and study-design limits with the result.
\end{enumerate}
The machine-readable schema remains in the appendix; the main text shows the claim, the family, a witness when answers differ, and the comparisons that remain fixed.

\section{What claim replay changes}

\subsection{Implications}

\textbf{Evaluation releases should expose a claim-replay interface, not only task code and a headline metric.}  For each advertised claim, the release should bind the historical observations or sufficient statistic, evaluator/judge version, support and missingness decisions, aggregation, admitted evaluator family, and the claim being asked.  The interface then either returns the supported answers or stops explicitly and names what is missing.

The required evidence depends on the claim.  A direction may be recoverable from a compact sufficient statistic, whereas an exact value can require sample-keyed outputs; a winner can be fixed even when the complete order varies.  Reporting \(\I_q(D,\Fpri)\) beside \(\I_q(D,\Frev)\), together with witnesses and stable pairs, makes sensitivity to disputed evaluator meanings reviewable.  Appendix~\ref{app:expanded-interface} gives the expanded release design.

\paragraph{Prospective utility remains an open test.}
This retrospective census does not estimate how many of the 110 stops a new release process would prevent.  A prospective test should register new units and advertised claims before outcomes, require the claim-replay fields at release time, and compare replay completion, time-to-audit, and decision changes against the legacy interface; Appendix~\ref{app:prospective-utility} gives the full design.

\subsection{Scope}

All headline results remain local to the named finite frame, evidence stream, frozen evidence, and enumerated evaluator family.  Appendices~\ref{app:technical-formalism}, \ref{app:frame-accounting}, and~\ref{app:reproducibility} give the full prevalence, coverage, independence, and transport boundaries.

\subsection{Related work}

\paragraph{Sensitivity analyses.}
Multiverse and specification-curve methods enumerate defensible analytic choices \citep{Steegen2016Multiverse,Simonsohn2020SpecificationCurve,Silberzahn2018ManyAnalysts}; benchmark work studies sensitivity to metrics, data, and protocols \citep{Flach2019PerformanceEvaluation,Bouthillier2021Variance,Dehghani2021BenchmarkLottery,DAmour2022Underspecification}.  Multicurious separates defensibility from equivalence \citep{ShortEtAl2026Multicurious}, matching our distinction between a proposed candidate and one that addresses the same target.  We retain cases where the evidence is unavailable, separate primary from disputed meanings, and report the answer for each claim.

\paragraph{Partial identification.}
Set-valued identification, fine-to-coarse functionals, and the separation of identification from sampling precision are established \citep{Tamer2010PartialIdentification,RomanoShaikh2008IdentifiableParameters,Kasy2016WelfareRanking}.  Robust ordinal regression similarly distinguishes necessary from possible preference relations \citep{GrecoEtAl2008OrdinalRegression}, and label-uncertainty work bounds performance under ambiguous labels \citep{PoloEtAl2024WeakSupervisionPI}.  We use this standard machinery for admitted evaluator families and add concrete disagreement witnesses and stable pairwise relations.

\paragraph{Benchmark infrastructure and measurement.}
Evaluation frameworks organize tasks, scenarios, metrics, and execution \citep{Liang2022HELM,Kiela2021Dynabench,Ribeiro2020CheckList}, while audit work stresses documentation and the limits of aggregate leaderboards \citep{Raji2021EverythingBenchmark,Bowman2021FixBenchmarking,Ethayarajh2020UtilityLeaderboards}.  Measurement research separates constructs from operationalizations \citep{Jacobs2021MeasurementFairness,Bean2025ConstructValidity,SalaudeenEtAl2025MeasurementMeaning}; judge diagnostics study evaluator competence and spurious gains \citep{ChenEtAl2026JudgeCompetence,XuEtAl2026PhantomGains,MahmudEtAl2026JaggedFrontier}.  We connect these layers through three direct questions: is the historical evidence sufficient, which evaluator meanings count, and what conclusions remain fixed?

\paragraph{Deployment-complete benchmarking.}
Mansouri and Arai \citeyearpar{MansouriArai2026DeploymentComplete}, the closest neighbor, ask what actions benchmark evidence licenses; we ask what conclusions historical evidence supports across admitted evaluator meanings.

\paragraph{Closest axes.}
Qin and Tong \citeyearpar{QinTong2026Observability} ask whether trajectories determine a diagnostic object under a fixed contract; Grynets et al.\ \citeyearpar{GrynetsEtAl2026SpecificationDetermines} vary conforming implementations of one specification; Frame-Level AUC changes comparison populations while holding scores fixed \citep{Abdulaziz2026FrameAUC}.  We ask the complementary historical question: given a pinned release and a stated claim, which answers are supported across the evaluator meanings admitted to review?

\noindent\textbf{Conclusion.}
A benchmark can be runnable while its historical claim is not replayable.  Claim replay states what evidence is missing and, when the evidence is sufficient, which answers change and which remain fixed across admitted evaluator meanings.  This turns ambiguity into a reviewable scientific result rather than one generic robustness label.

\clearpage
\subsection*{AI use statement}

In this work, generative AI tools, principally OpenAI Codex, assisted with refining and operationalizing the conceptual framework and methodology; formulating and checking mathematical claims; refining hypotheses; implementing and testing code; cleaning and reformatting bound data; supporting qualitative audits and checking candidate interpretations of results against bound evidence; translation; creating and editing figures; searching and summarizing literature; and drafting, editing, and formatting the paper and reproducibility artifacts.  They were not used to generate synthetic data or collect new experimental observations.  No AI system was assigned final authority over claim selection, semantic admission, evidence licensing, interpretation, or reporting.  The authors reviewed the AI-assisted text, code, data transformations, citations, figures, and results.  Numerical and code outputs were checked with deterministic validators and, where reported, independent recomputations from bound source tables; cited claims were checked against primary sources.  The authors made all final scientific and reporting decisions and take responsibility for all contents, claims, and artifacts.

\clearpage
\bibliography{references}

\fontsize{9.0}{10.20}\selectfont
\clearpage
\appendix
\raggedbottom
\small

\section{Claim hierarchy, typing, and release interface}
\label{app:reproducibility}

The three stages in Figure~\ref{fig:missing-layer} answer different questions.  Evidence stopping is an \emph{executability state}: the historical observations, judge traces, terminal state, support joins, or missingness decisions needed by a claim are unavailable.  Semantic multiplicity is a \emph{family property}: more than one grounded reading survives type, support, reliability, and admission review.  Claim-resolution separation is an \emph{inference result}: after deterministic enumeration, exact values, winners, complete orders, and pairwise relations can have different identified sets.  Treating all three as generic sensitivity outcomes would condition away unavailable cases and discard stable coarsenings.

\subsection{Technical evaluator map}
\label{app:technical-formalism}

A grounded specification \(f\) can be expanded into a semantic map \(M_f\), row rule \(\ell_f\), and aggregation rule \(A_f\):
\begin{equation}
  H_i(f),e_i(f)=M_f(x_i),\qquad
  Y_{im}(f)=\ell_f(p_{im},H_i(f)),\qquad
  V_m(f;D)=A_f\{Y_{im}(f):e_i(f)=1\}.
  \label{eq:appendix-spec-map}
\end{equation}
Here \(H_i(f)\) is the specification-dependent target, \(e_i(f)\) its eligibility indicator, and \(Y_{im}(f)\) the row-level value for system \(m\).  The decomposition forces eligibility, row scoring, and aggregation to be declared separately.  The endpoint is typed as a scalar, paired contrast, direction, top set, or weak order.  Exact-value identification is formed only for commensurate scalar endpoints; ordinal families may display per-specification values without pooling their magnitudes.

Semantic and sampling variation are different coordinates.  A specification image varies \(f\) on fixed \(D\); a bootstrap or confidence interval varies observations under one fixed \(f\).  Published-table precision is a third coordinate.  Reversible unit changes are a fourth: if \(V'=aV+b\) with \(a>0\), winners, orders, and pair relations are invariant even though printed values differ.  These distinctions prevent a representation change from becoming false semantic multiplicity.

\subsection{Release-interface implications}
\label{app:expanded-interface}

A minimal claim-replay release binds: (i) system and observation identifiers; (ii) evaluator or judge version and configuration; (iii) support and missingness masks; (iv) a typed aggregation tree; (v) candidate origin, reliability screen, and admission; (vi) a pre-outcome family-freeze receipt; and (vii) a deterministic claim ledger.  Stateful benchmarks may expose sufficient state manifests and checker outputs instead of full images when that closes the declared query.  Judge-based benchmarks need sample-keyed judge decisions and prompt/model bindings, not only judge code.

The required object is claim-relative rather than uniformly maximal.  A direction may be recoverable from a compact sufficient statistic, while exact-value replay can require sample-keyed outputs; a winner can be identified even when the complete order is not.  The contract therefore asks for the least artifact that closes the declared query, together with a machine-checkable sufficiency relation.  It does not certify a uniquely correct evaluator meaning or a complete semantic family.

\begin{figure}[ht]
\centering
\resizebox{.96\linewidth}{!}{%
\begin{tikzpicture}[
  font=\scriptsize,
  >=Latex,
  axis/.style={draw,rounded corners=2pt,align=center,text width=3.25cm,minimum height=1.05cm,inner sep=4pt},
  outbox/.style={draw,rounded corners=2pt,align=center,text width=4.05cm,minimum height=.92cm,inner sep=4pt},
  arrow/.style={->,line width=.65pt}
]
\node[axis,fill=orange!10] (evidence) at (0,2.0) {\textbf{Evidence coordinate}\\Is historical \(D\) executable?};
\node[axis,fill=green!9] (semantic) at (4.2,2.0) {\textbf{Semantic coordinate}\\Which grounded \(f\) enters each scope?};
\node[axis,fill=blue!8] (sampling) at (8.4,2.0) {\textbf{Sampling coordinate}\\How uncertain is one fixed \(f\)?};
\node[axis,fill=gray!10] (display) at (12.6,2.0) {\textbf{Representation coordinate}\\Precision, units, reversible maps};
\node[outbox,fill=purple!8] (claim) at (4.2,0) {\textbf{Claim image}\\value, direction, winner, order};
\node[outbox,fill=purple!5] (structure) at (9.0,0) {\textbf{Recovered structure}\\minimum witness + stable pairs};
\draw[arrow] (evidence) -- (claim);
\draw[arrow] (semantic) -- (claim);
\draw[arrow] (sampling) -- (structure);
\draw[arrow] (display) -- (structure);
\draw[arrow] (claim) -- (structure);
\end{tikzpicture}%
}
\caption{\textbf{Four coordinates that a scalar sensitivity range can conflate.} The present contract enumerates semantic variation on fixed evidence, keeps sampling uncertainty separate, and removes purely reversible representation changes before claim comparison.}
\label{fig:appendix-four-coordinates}
\end{figure}
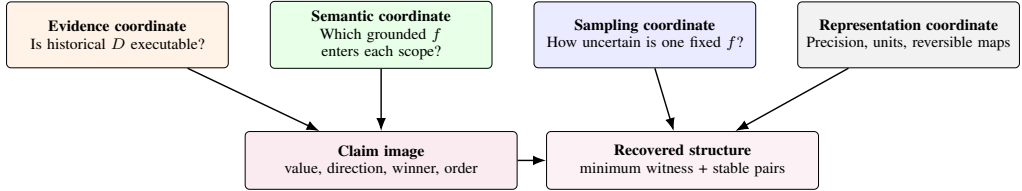

\subsection{Scientific validation boundary}

\begin{table}[ht]
\centering
\caption{Two validation layers and their distinct licenses.}
\label{tab:validation-boundary}
\footnotesize
\begin{tabularx}{\linewidth}{@{}p{.20\linewidth}p{.34\linewidth}Y@{}}
\toprule Layer & What is checked & Licensed conclusion \\
\midrule
Mechanical closure & Hashes and schema; terminal accounting; deterministic source-table recomputation; family projection; claim images, witnesses, and stable pairs & The bound bytes produce the reported arithmetic and identified sets through an independently checked code path \\
Human-auditable semantic provenance & Source locators and quotations; normalized authorization; same-target/support rationale; A/D/X judgments; evidence horizon & An independent reader can inspect, dispute, and correct the semantic ledger; the automated verifier does not prove construct validity or provenance correctness \\
Outside both layers & Unreleased sample-level facts; unenumerated semantics; future releases or interventions & No family-completeness, ecosystem-prevalence, independent-replication, benchmark-validity, causal-utility, or future-stop-reduction claim \\
\bottomrule
\end{tabularx}
\end{table}

\section{Finite frame and typed stopping}
\label{app:frame-accounting}

\subsection{Mechanical frame construction}
\label{app:frame-predicate}

At public commit \texttt{32b79a2}, the frozen tree contains 129 paths matching exactly \texttt{src/inspect\_evals/<id>/eval.yaml}.  Each file must parse as a YAML mapping; the predeclared stratum mapping is applied mechanically; and every required external asset must be marked pinned or controlled.  Exactly five units are excluded: one prior audit and four packages with at least one floating required asset.  Thus \(129-5=124\).  No outcome enters this predicate.

\begin{table}[ht]
\centering
\caption{Nested accounting for the commit-bound frame.  The rows are accounting views, not pooled result samples.}
\label{tab:reviewer-frame-accounting}
\begin{tabularx}{\linewidth}{@{}p{0.23\linewidth}p{0.16\linewidth}p{0.14\linewidth}Y@{}}
\toprule View & Complete & \STOP{} & Licensed interpretation \\
\midrule
All 124 & 14 & 110 & Terminal auditability inside the complete eligible frame \\
Outcome-blind census-review path (114) & 11 & 103 & Source review, admission, and fixed-observation execution before outcome inspection \\
Frozen random ten & 3 & 7 & Completion under the balanced two-per-stratum draw; stopped units retained \\
\bottomrule
\end{tabularx}
\end{table}

\begin{minipage}{\linewidth}
\noindent\textbf{Object-level exclusions.}  The five manifest records make the subtraction auditable rather than nominal.  \texttt{agentdojo} carries \texttt{previously\_audited}, a deduplication exclusion rather than an asset failure.  \texttt{gdm\_in\_house\_ctf} pulls per-challenge \texttt{marshw/*} Docker images at runtime without immutable digests.  \texttt{osworld} pins its repository revision but also downloads an unbound \texttt{fonts.tar.gz} from a moving branch.  The \texttt{piqa} builder fetches an unbound training/development ZIP and test JSONL even though its Hugging Face snapshot is pinned.  \texttt{swe\_lancer} pulls per-issue and monolith images under the mutable \texttt{releasev1} tag rather than immutable digests.  A single floating required asset triggers exclusion; the label does not assert that the benchmark is defective or cannot currently run.
\end{minipage}

\begin{figure}[ht]
\centering
\resizebox{\linewidth}{!}{%
\begin{tikzpicture}[>=Latex,line width=.62pt,font=\small]
  \draw[gray!55] (-.55,-1.38) rectangle (15.7,1.02);
  \node[font=\Large\bfseries] (n129) at (0.15,.32) {129};
  \node[font=\Large\bfseries] (n124) at (4.50,.32) {124};
  \node[font=\normalsize\bfseries,anchor=west] (terminal) at (8.1,.32) {14 COMPLETE $+$ 110 TYPED STOPS};
  \draw[->] (n129.east) -- node[above=4pt,font=\bfseries] {eligibility}
    node[below=4pt,align=center] {1 prior audit\\4 floating assets} (n124.west);
  \draw[->] (n124.east) -- node[above=4pt,font=\bfseries] {zero substitution} (terminal.west);
  \draw[gray!55,dashed] (-.30,-.42) -- (15.45,-.42);
  \node[anchor=west,font=\bfseries] at (-.15,-.95) {Evidence panels (never pooled):};
  \node[draw,rounded corners=1pt,inner xsep=6pt,inner ysep=3pt] at (7.65,-.95) {3 random complete};
  \node[draw,rounded corners=1pt,inner xsep=6pt,inner ysep=3pt] at (11.15,-.95) {3 purposive deep};
  \node[draw,rounded corners=1pt,inner xsep=6pt,inner ysep=3pt] at (14.25,-.95) {8 exploratory};
\end{tikzpicture}%
}
\caption{\textbf{Two accounting axes.} The top line exhausts the 124-unit frame.  The bottom line names evidence panels with distinct selection and exposure licenses; arithmetic equality does not make them one cohort.}
\label{fig:reviewer-frame-accounting}
\end{figure}
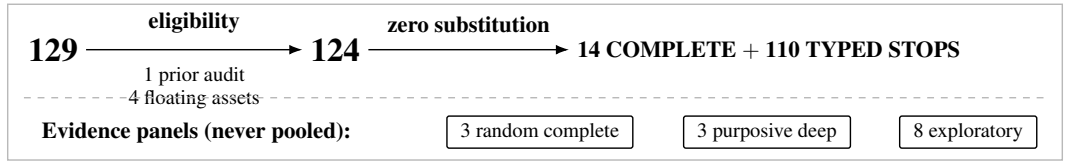

The random allocation is disproportionate because stratum sizes \((47,40,26,7,4)\) are unequal while each contributes two units.  Its (3/10) is the observed completion fraction in that frozen balanced-strata draw and targets an equal-stratum-weighted completion estimand under within-stratum random sampling.  It is not a unit-weighted frame rate and not a rate of non-identification: seven selected units stop before any claim image exists.

\subsection{Native logging and the claim-replay boundary}
\label{app:native-log-binding}

The audited snapshot contains substantial native logging capacity.  A retained \texttt{EvalLog} can carry sample inputs, outputs, targets, scores, metadata, and reductions.  This positive capability does not imply that every historical log was published, or that the log declares alternative semantics, same-target/support admission, a typed claim query, identified sets, witnesses, or a stable backbone.  The added layer binds that final claim-relative bridge.

\subsection{Preservation layers}
\label{app:preservation-layers}

\begin{figure}[ht]
\centering
\resizebox{\linewidth}{!}{%
\begin{tikzpicture}[
  font=\scriptsize,
  >=Latex,
  layer/.style={draw,rounded corners=2pt,align=center,text width=2.35cm,minimum height=1.02cm,inner sep=3pt},
  stop/.style={align=center,text width=2.35cm,font=\tiny},
  arrow/.style={->,line width=.65pt}
]
\node[layer,fill=blue!8] (source) at (0,1.25) {\textbf{Source}\\task + scorer};
\node[layer,fill=blue!8] (obs) at (3.05,1.25) {\textbf{Observation}\\fixed outputs};
\node[layer,fill=blue!8] (state) at (6.10,1.25) {\textbf{Mediator/state}\\judge or terminal state};
\node[layer,fill=blue!8] (support) at (9.15,1.25) {\textbf{Support}\\keys + DNF rules};
\node[layer,fill=green!9] (family) at (12.20,1.25) {\textbf{Semantic registry}\\type + A/D/X + query};
\draw[arrow] (source) -- (obs);
\draw[arrow] (obs) -- (state);
\draw[arrow] (state) -- (support);
\draw[arrow] (support) -- (family);
\node[stop] at (0,0) {source/evaluator\\binding stop};
\node[stop] at (3.05,0) {comparative\\observation stop};
\node[stop] at (6.10,0) {judge/state\\stop};
\node[stop] at (9.15,0) {support or\\missingness stop};
\node[stop] at (12.20,0) {typing, target,\\or claim stop};
\end{tikzpicture}%
}
\caption{\textbf{The preservation chain is claim-relative.} A retained scorer is only the first layer; later claims can require historical outputs, mediator decisions, state, support joins, and an admitted semantic registry.  The preservation layers are conceptual dependencies; the verifier precedence used for first-stop accounting is G1--G8 below.}
\label{fig:reviewer-preservation-chain}
\end{figure}
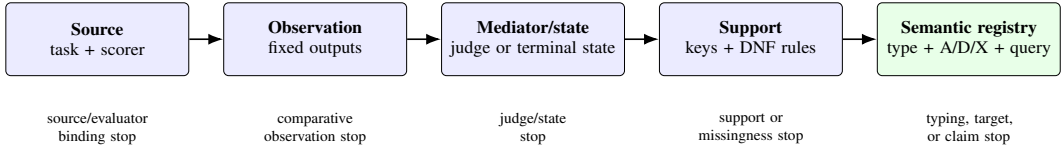

\begin{table}[ht]
\centering
\caption{Preservation layers required by claim replay.}
\label{tab:reviewer-preservation-layers}
\footnotesize
\begin{tabularx}{\linewidth}{@{}p{0.16\linewidth}p{0.29\linewidth}p{0.25\linewidth}Y@{}}
\toprule Layer & Required object & Enables & Typed stop when absent \\
\midrule
Source & task, scorer, registry revision, dependencies & Ground candidate formulas & source/evaluator binding \\
Observation & sample-keyed outputs and system/prompt binding & Apply alternatives to the same responses & comparative observations \\
Mediator/state & judge decisions or sufficient terminal state & Replay mediated or stateful endpoints & judge trace or execution state \\
Support & keys, DNF/missingness masks, denominators & Compare without silent eligibility change & support or missingness \\
Semantic registry & endpoint type, origin/admission, same target/support, units/ties & Form \((D,\F,q)\) & family or claim review \\
\bottomrule
\end{tabularx}
\end{table}

\begin{table}[ht]
\centering
\caption{Primary stopping codes for the 103 stops on the outcome-blind 114-case path.}
\label{tab:reviewer-stop-taxonomy}
\footnotesize
\begin{tabularx}{\linewidth}{@{}p{0.29\linewidth}rY@{}}
\toprule Reader label & Count & First unavailable requirement \\
\midrule
Evidence binding & 47 & Source or observation bytes could not be jointly bound \\
Comparative observations & 35 & Fixed outputs for a common system/support set were absent \\
Judge trace & 9 & Responses, judge binding, and decisions were not jointly frozen \\
Endpoint type & 4 & Available outputs lacked a shared typed endpoint \\
Semantic target & 4 & Candidate metrics addressed different targets \\
Missingness / DNF & 2 & Missing outcomes lacked a declared non-ignorable rule \\
Claim selection & 1 & No single claim query could be bound without further judgment \\
Execution state & 1 & A stateful evaluator required unreleased terminal state \\
\bottomrule
\end{tabularx}
\end{table}

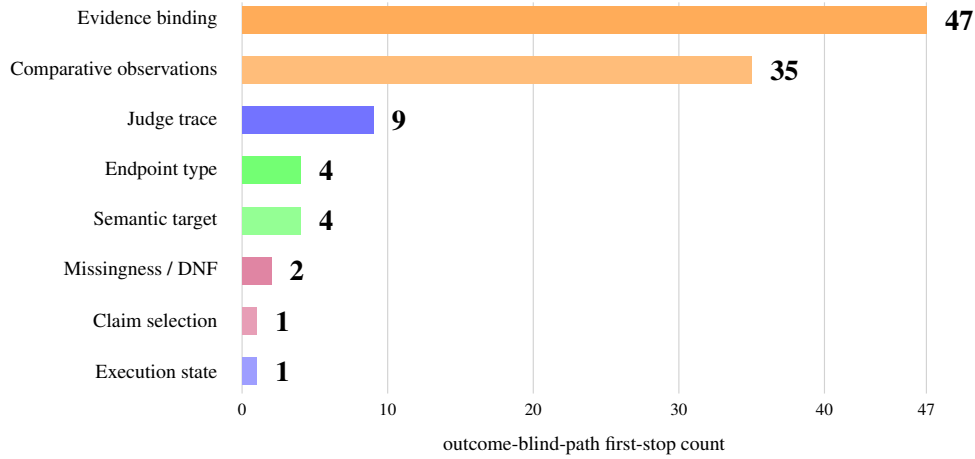
\begin{figure}[ht]
\centering
\resizebox{.93\linewidth}{!}{%
\begin{tikzpicture}[font=\scriptsize,x=.18cm,y=.62cm]
  \foreach \x/\lab in {0/0,10/10,20/20,30/30,40/40,47/47} {
    \draw[gray!35] (\x,.35) -- (\x,-7.45);
    \node[below,font=\tiny] at (\x,-7.45) {\lab};
  }
  \foreach \name/\count/\y/\barcolor in {
    {Evidence binding}/47/0/orange!65,
    {Comparative observations}/35/-1/orange!52,
    {Judge trace}/9/-2/blue!55,
    {Endpoint type}/4/-3/green!55,
    {Semantic target}/4/-4/green!42,
    {Missingness / DNF}/2/-5/purple!48,
    {Claim selection}/1/-6/purple!38,
    {Execution state}/1/-7/blue!38}
  {
    \node[anchor=east] at (-1,\y) {\name};
    \fill[\barcolor] (0,\y-.27) rectangle (\count,\y+.27);
    \node[anchor=west,font=\bfseries] at (\count+.6,\y) {\count};
  }
  \node[below,font=\scriptsize] at (23.5,-8.1) {outcome-blind-path first-stop count};
\end{tikzpicture}%
}
\caption{\textbf{First unavailable requirement among the 103 stops on the outcome-blind 114-case path.} The distribution identifies release-interface bottlenecks; it is not a benchmark-quality score and not evidence of intervention utility.}
\label{fig:reviewer-stop-bars}
\end{figure}

The primary stop is the first failed gate in a fixed order: G1 source binding, G2 claim selection, G3 endpoint typing, G4 observation binding, G5 evaluator executability, G6 family adjudication, G7 deterministic execution, and G8 synthesis/verification; registration precedes G1.  Simultaneously missing requirements remain secondary modes.  First-stop totals are 47 at source binding, one at claim selection, eight at endpoint typing, and 47 at observation binding.  Each terminal record names the first unavailable object, secondary modes, no-imputation reason, fresh-run boundary, and minimum unblock.  The (110) stops are therefore a preservation diagnosis, not a zero-valued performance result or evidence that the benchmarks are defective.

\subsection{Search surface and fail-closed boundary}
\label{app:stop-search-boundary}

For each unit, the systematic public search surface comprises the pinned repository subtree and files referenced by its registry entry; the declared paper, README, and citation links; official project documentation; and evaluator, dataset, or source assets linked from those surfaces.  Searches are evaluated at the recorded evidence horizon and end only after these surfaces have been inspected for the artifact required by the typed query.  If the needed historical bytes cannot be jointly bound to system, support, evaluator, and version, the record fails closed at the first unmet gate and names the minimum unblock.  This is not a claim that no artifact exists anywhere on the web or in private storage.  A later demonstration that a public, version-bindable artifact inside the declared surface was missed is a corrigible provenance error and should enter as a visible amendment rather than be converted silently into a score.  The seven random-ten stopping records below expose representative object-level trails; the reproduction package retains the terminal record and locator fields for the complete frame.

\section{Complete audits and recovered structure}
\label{app:complete-audit-records}

The random and purposive panels have different acquisition licenses and are not pooled.  Completion means every admitted member of the declared finite family was evaluated on the frozen substrate; it does not certify family completeness.

\begin{figure}[ht]
\centering
\begin{tikzpicture}[font=\scriptsize,
  idbox/.style={draw,rounded corners=1pt,minimum width=1.18cm,minimum height=.43cm,fill=green!13},
  nidbox/.style={draw,rounded corners=1pt,minimum width=1.18cm,minimum height=.43cm,fill=orange!13},
  stablebox/.style={draw,rounded corners=1pt,minimum width=1.62cm,minimum height=.43cm,fill=blue!8}]
  \node[anchor=east,font=\bfseries] at (3.0,.72) {Case};
  \node[font=\bfseries] at (4.35,.72) {P winner};
  \node[font=\bfseries] at (5.95,.72) {P order};
  \node[font=\bfseries] at (7.55,.72) {E winner};
  \node[font=\bfseries] at (9.15,.72) {E order};
  \node[font=\bfseries] at (11.25,.72) {stable P/E};
  \draw[gray!45] (.55,.40) -- (12.15,.40);

  \foreach \name/\y in {{TheAgentCompany}/0,{BBEH}/-.65,{Personality}/-1.30,{AgentDojo}/-2.15,{AutoML}/-2.80,{BEIR SciFact}/-3.45} {
    \node[anchor=east] at (3.0,\y) {\name};
  }
  \foreach \y in {0,-.65,-1.30} {
    \node[nidbox] at (4.35,\y) {NID}; \node[nidbox] at (5.95,\y) {NID};
    \node[nidbox] at (7.55,\y) {NID}; \node[nidbox] at (9.15,\y) {NID};
  }
  \node[stablebox] at (11.25,0) {0/1; 0/1};
  \node[stablebox] at (11.25,-.65) {0/1; 0/1};
  \node[stablebox] at (11.25,-1.30) {0/3; 0/3};

  \node[idbox] at (4.35,-2.15) {ID}; \node[idbox] at (5.95,-2.15) {ID};
  \node[idbox] at (7.55,-2.15) {ID}; \node[nidbox] at (9.15,-2.15) {NID};
  \node[stablebox] at (11.25,-2.15) {3/3; 2/3};
  \node[idbox] at (4.35,-2.80) {ID}; \node[idbox] at (5.95,-2.80) {ID};
  \node[idbox] at (7.55,-2.80) {ID}; \node[nidbox] at (9.15,-2.80) {NID};
  \node[stablebox] at (11.25,-2.80) {10/10; 9/10};
  \node[idbox] at (4.35,-3.45) {ID}; \node[nidbox] at (5.95,-3.45) {NID};
  \node[idbox] at (7.55,-3.45) {ID}; \node[nidbox] at (9.15,-3.45) {NID};
  \node[stablebox] at (11.25,-3.45) {2/3; 2/3};

  \draw[gray!35] (.55,-1.72) -- (12.15,-1.72);
  \draw[blue!55,line width=2.5pt] (.75,.18) -- (.75,-1.48);
  \node[rotate=90,font=\tiny,text=blue!55!black] at (.27,-.65) {random complete};
  \draw[green!55!black,line width=2.5pt] (.75,-1.92) -- (.75,-3.68);
  \node[rotate=90,font=\tiny,text=green!45!black] at (.27,-2.80) {purposive deep};
\end{tikzpicture}
\caption{\textbf{Claim resolution across the six complete audits.} P/E denote primary/review scopes.  The last column reports stable pair counts.  AgentDojo and AutoML make the admission effect visible: their primary orders are identified even though their review-envelope orders are not.  The BEIR row is scoped to its declared five-view family; the separate 25-view coverage sensitivity is reported below.}
\label{fig:reviewer-complete-resolution}
\end{figure}

\clearpage
\Needspace{19\baselineskip}
\subsection{How much pairwise structure survives?}

Complete-order non-identification is not the endpoint of the audit.  The stable-backbone count asks how many system pairs keep the same relation across the declared family.  The random completions genuinely have no stable pair, whereas the three purposive audits retain an identified winner under their declared families and two retain most lower relations even in the review envelope.

\begin{figure}[ht]
\centering
\resizebox{.99\linewidth}{!}{%
\begin{tikzpicture}[font=\scriptsize,x=.10cm,y=.90cm]
  \foreach \x/\lab in {0/0,25/25,50/50,75/75,100/100} {
    \draw[gray!30] (\x,.65) -- (\x,-5.55);
    \node[above,font=\tiny] at (\x,.65) {\lab\%};
  }
  \foreach \name/\p/\e/\labelp/\labele/\y in {
    {TheAgentCompany}/0/0/{0/1}/{0/1}/0,
    {BBEH}/0/0/{0/1}/{0/1}/-1,
    {Personality}/0/0/{0/3}/{0/3}/-2,
    {AgentDojo}/100/66.7/{3/3}/{2/3}/-3,
    {AutoML}/100/90/{10/10}/{9/10}/-4,
    {BEIR SciFact}/66.7/66.7/{2/3}/{2/3}/-5}
  {
    \node[anchor=east] at (-3,\y) {\name};
    \fill[blue!55] (0,\y+.04) rectangle (\p,\y+.26);
    \fill[green!55!black] (0,\y-.26) rectangle (\e,\y-.04);
    \node[anchor=west,font=\tiny] at (\p+1.2,\y+.15) {P \labelp};
    \node[anchor=west,font=\tiny] at (\e+1.2,\y-.15) {E \labele};
  }
  \foreach \y in {0,-1,-2} {
    \draw[blue!65!black,fill=white,line width=.7pt] (0,\y+.15) circle (1.8pt);
    \draw[green!55!black,fill=white,line width=.7pt] (0,\y-.15) circle (1.8pt);
  }
  \node[anchor=west,font=\tiny,text=blue!65!black] at (0,-6.1) {P = primary};
  \node[anchor=west,font=\tiny,text=green!45!black] at (28,-6.1) {E = review envelope};
\end{tikzpicture}%
}
\caption{\textbf{Stable-backbone retention across declared complete-audit families.} Bar length is the fraction of system pairs whose relation is invariant; labels give exact stable/total counts.  The BEIR bar is 2/3 over the declared five views, not an evaluator-wide claim: its 25-view sensitivity yields 0/3.}
\label{fig:reviewer-stable-backbone}
\end{figure}
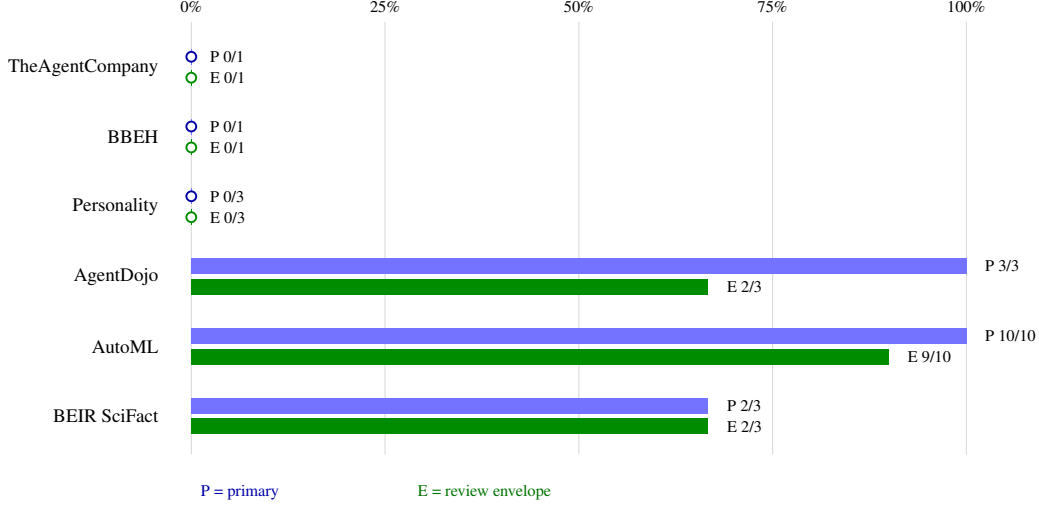

\begin{table}[ht]
\centering
\caption{Provenance correction and family expansion are different perturbations.}
\label{tab:automl-beir-contrast}
\footnotesize
\begin{tabularx}{\linewidth}{@{}p{.13\linewidth}p{.31\linewidth}p{.25\linewidth}Y@{}}
\toprule Case & Audit intervention & Claim effect & Licensed lesson \\
\midrule
AutoML & Reclassify average task rank from AP/D to SM/A after binding the canonical missing-value, fold, task-rank, and average-rank procedure & Ledger changes from 3A+2D to 4A+1D; winner, order image, 10/10 primary pairs, and 9/10 envelope pairs are unchanged & Provenance can be corrected without destabilizing inference \\
BEIR SciFact & Retain the declared five views, then separately expand to all 25 evaluator-exposed metric--cutoff cells executable from the top-100 runs & Winner ID\(\to\)NID; order remains NID but expands to five weak orders; backbone 2/3\(\to\)0/3 & Mechanically correct arithmetic does not license transport beyond the named family \\
\bottomrule
\end{tabularx}
\end{table}

\begin{table}[ht]
\centering
\caption{One-at-a-time boundary perturbations propagated through the claim calculation.}
\label{tab:admission-perturbations}
\footnotesize
\begin{tabularx}{\linewidth}{@{}p{.15\linewidth}p{.31\linewidth}p{.18\linewidth}Y@{}}
\toprule Case & Perturbation & Winner after perturbation & Complete-order consequence \\
\midrule
AgentDojo & common-key suite-macro ASR, D\(\to\)A alone & Claude 3.5 Sonnet & ID; primary order unchanged \\
AgentDojo & utility-conditional ASR, D\(\to\)A alone & Claude 3.5 Sonnet & NID; GPT-4o-mini/Gemini pair reverses \\
AgentDojo & worst-suite ASR, D\(\to\)A alone & Claude 3.5 Sonnet & NID; GPT-4o-mini/Gemini pair reverses \\
AutoML & worst-task AUC, D\(\to\)A alone & H2OAutoML & NID; AutoWEKA/TPOT pair reverses \\
BEIR SciFact & No D candidate; five\(\to\)25 is a coverage, not admission, perturbation & Non-identified & NID; all three pairs become unstable \\
\bottomrule
\end{tabularx}
\end{table}

These rows make the semantic disagreement executable rather than rhetorical.  They do not declare every D\(\to\)A move reasonable; they show exactly what each already-recorded boundary judgment would change.

\subsection{Controlled contract and robustness micro-experiments}
\label{app:microexperiments}

These deterministic micro-experiments reuse already-bound records; they add no benchmark run, model or judge call, annotation, or claim type.  Random-complete and purposive cases retain their separate design labels.  The frozen companion artifact rebuilds the three JSON summaries from portable hash-bound inputs and then runs an independent verifier (844/844 checks).

\paragraph{Controlled evidence-ablation conformance.}
Five prospective \texttt{COMPLETE\_DETERMINISTIC} packets were replayed intact and after deleting exactly one required evidence class.  The intact controls remained complete, and every deletion stopped at its declared boundary (Table~\ref{tab:microexp-ablation}).  In particular, downstream copies were not used to repair the removed evidence.  This is fail-closed contract conformance: it establishes that the executable implementation follows the declared stopping contract under controlled evidence deletion, not that the gates are causally or semantically correct.

\begin{table}[ht]
\centering
\caption{Controlled evidence-ablation conformance.  Cells report the replay terminal after the named single-class deletion.}
\label{tab:microexp-ablation}
\footnotesize
\setlength{\tabcolsep}{3pt}
\begin{tabularx}{\linewidth}{@{}p{.21\linewidth}ccccc@{}}
\toprule Packet & Intact & Source & Claim & Endpoint & Observations \\
\midrule
ComputeEval (021) & COMPLETE & G1 STOP & G2 STOP & G3 STOP & G4 STOP \\
MLRC-Bench (043) & COMPLETE & G1 STOP & G2 STOP & G3 STOP & G4 STOP \\
TAC (086) & COMPLETE & G1 STOP & G2 STOP & G3 STOP & G4 STOP \\
UCCB (101) & COMPLETE & G1 STOP & G2 STOP & G3 STOP & G4 STOP \\
Pre-Flight (105) & COMPLETE & G1 STOP & G2 STOP & G3 STOP & G4 STOP \\
\bottomrule
\end{tabularx}
\end{table}

\paragraph{Family-boundary stress.}
Table~\ref{tab:microexp-family-stress} recomputes each typed resolution over \(\Fpri\) and \(\Frev\).  Two complete orders lose identification, no previously identified winner does, and 13/15 previously identified pairwise relations remain identified.  The remaining six pairwise relations were already non-identified in the primary scope, leaving 13/21 stable in review.  The declared initial union equals the review family in all six audits; it is therefore a structural zero, not a third robustness level.

\begin{table}[ht]
\centering
\caption{Claim-relative family stress. P/E denote primary/review; pair entries are stable/total.}
\label{tab:microexp-family-stress}
\footnotesize
\setlength{\tabcolsep}{3pt}
\begin{tabularx}{\linewidth}{@{}p{.22\linewidth}p{.18\linewidth}ccc@{}}
\toprule Case & Design & Winner P/E & Order P/E & Pairs P/E \\
\midrule
TheAgentCompany & random complete & NID/NID & NID/NID & 0/1 $\to$ 0/1 \\
BBEH & random complete & NID/NID & NID/NID & 0/1 $\to$ 0/1 \\
Personality & random complete & NID/NID & NID/NID & 0/3 $\to$ 0/3 \\
AgentDojo & purposive & ID/ID & ID/NID & 3/3 $\to$ 2/3 \\
AutoML & purposive & ID/ID & ID/NID & 10/10 $\to$ 9/10 \\
BEIR SciFact & purposive & ID/ID & NID/NID & 2/3 $\to$ 2/3 \\
\bottomrule
\end{tabularx}
\end{table}

The separate BEIR 25-view \emph{exploratory coverage sensitivity} changes winner ID\(\to\)NID, leaves complete order NID, and contracts the stable backbone from 2/3 to 0/3.  It is visually and inferentially separate from the declared-family comparison: it does not enter the headline 13/15 denominator and does not certify the 25 views as an admitted family.

\paragraph{Disagreement witnesses.}
The six declared review families yield 31 distinct specification pairs witnessing at least one query disagreement after deduplicating repeated query uses: 14 admitted--admitted, 15 admitted--disputed, and two disputed--disputed.  Thus non-identification does not require admitting a disputed challenger.  These are the recorded admission categories; the result does not promote audit-proposed specifications to independently grounded semantics or certify family completeness.

\paragraph{Challenger-removal persistence.}
Across 17 non-identified query instances in those review families, 76 leave-one-out deletions were recomputed.  Disagreement persisted after 71 deletions, while five restored identification.  This asks whether the full non-identification is pivotal on a single specification; it is distinct from the pair-witness question above.  Cardinality two is expected for deterministic point-valued disagreement and is not itself a robustness claim.  The separately labelled BEIR 25-view diagnostic is excluded from these declared-family totals.

The complete records below retain all candidate definitions, exact outputs, witnesses, pair relations, residual missing material, seven zero-replacement stopping mechanisms, and three purposive audits.  Only object-level digests and filesystem-oriented provenance tables have been moved to the public reproduction package.

\clearpage
\begingroup
\setlength{\emergencystretch}{3em}
\section{Thirteen complete audit and stopping records}\label{app:r8-census}
The scientific evidence portfolio is partitioned before inference.  A frozen stratified draw contributes ten objects with zero substitution: three complete finite-family audits and seven infrastructure/evidence stopping records.  Three additional objects were selected purposively for deep, cross-domain mechanism audits.  Consequently, the only sample completion statement is \textbf{3/10}; the purposive cases are not added to that probability denominator.  A stopping record is a protocol outcome under captured public resources, not a benchmark defect or a silently imputed failure.
\begingroup\scriptsize
\begin{longtable}{@{}p{0.18\linewidth}p{0.25\linewidth}p{0.49\linewidth}@{}}
\toprule Cohort & Object & Controlling audit outcome \\
\midrule\endfirsthead
\toprule Cohort & Object & Controlling audit outcome \\
\midrule\endhead
Random 10 & GDM InterCode CTF & stopped after semantic candidates: comparative observation absent \\
Random 10 & TheAgentCompany & complete finite-family audit \\
Random 10 & FrontierScience & stopped after semantic candidates: model judge trace absent \\
Random 10 & BBEH & complete finite-family audit \\
Random 10 & Agent Bench & stopped after semantic candidates: stateful execution artifact absent \\
Random 10 & CORE-Bench & stopped after semantic candidates: stateful execution artifact absent \\
Random 10 & MMIU & stopped after semantic candidates: comparative observation absent \\
Random 10 & MMMU & stopped after semantic candidates: model judge trace absent \\
Random 10 & Personality & complete finite-family audit \\
Random 10 & BOLD & stopped after semantic candidates: endpoint target fragmentation \\
Purposive deep & AgentDojo & ordinal; P: winner/order yes, 3/3; E: winner yes, order no, 2/3 \\
Purposive deep & AutoML Benchmark & ordinal; P: winner/order yes, 10/10; E: winner yes, order no, 9/10 \\
Purposive deep & BEIR SciFact & ordinal; P=E: winner yes, order no, 2/3 \\
\bottomrule\end{longtable}\endgroup
\section{Frozen Random Ten: Three Complete Audits}\label{app:r8-random-complete}
Each subsection gives the declared family, typed endpoint, exact reconstructed outputs, winner/order identified sets, minimal witness, pairwise stability, remaining missing material.  Completion means every admitted member of the finite family was evaluated on the frozen summary substrate; it does not certify natural or semantic-family exhaustion.

\subsection{TheAgentCompany: complete finite-family audit}\label{app:r8-theagentcompany}

\paragraph{Typed record.} Query: Is the ordering of the two documented runs stable between checkpoint progress and strict task completion? Endpoint: ordinal ranking of two fixed 34-task reported vectors. Fixed substrate (frozen\_summary\_vectors): 34 rounded task rows in the frozen README, with system and grader identities documented. The audit evaluates every admitted member of the declared finite family; it does not certify family completeness.

\begingroup
\scriptsize
\begin{tabularx}{\linewidth}{@{}p{0.27\linewidth}p{0.14\linewidth}p{0.18\linewidth}X@{}}
\toprule
Specification & Admission & Origin & Semantic coordinate \\
\midrule
mean-checkpoint-progress & admitted & source-documented & partial credit \\
strict-task-completion & admitted & source-mandated & partial credit versus binary completion \\
any-progress & disputed & audit-proposed & eligibility threshold and partial credit \\
\bottomrule
\end{tabularx}
\endgroup

\paragraph{Exact outputs.} No additional rounding of the reconstructed values is used to determine the claims.

\begingroup
\scriptsize
\begin{longtable}{@{}p{0.25\linewidth}p{0.39\linewidth}p{0.27\linewidth}@{}}
\toprule
System/artifact & Specification output & Exact reconstructed value \\
\midrule
\endfirsthead
\toprule System/artifact & Specification output & Exact reconstructed value \\
\midrule
\endhead
inspect\_haiku45 & any\_progress & \texttt{0.6176470588235294} \\
inspect\_haiku45 & mean\_checkpoint & \texttt{0.36264705882352943} \\
inspect\_haiku45 & strict\_completion & \texttt{0.11764705882352941} \\
original\_sonnet4 & any\_progress & \texttt{0.47058823529411764} \\
original\_sonnet4 & mean\_checkpoint & \texttt{0.2164705882352941} \\
original\_sonnet4 & strict\_completion & \texttt{0.14705882352941177} \\
\bottomrule
\end{longtable}
\endgroup

\paragraph{Primary and review-envelope identified sets.} Primary (2A): winner and complete order are non-identified; the winner image is \{inspect\_haiku45\}; \{original\_sonnet4\}, and the order image is inspect\_haiku45 > original\_sonnet4; original\_sonnet4 > inspect\_haiku45. Review envelope (2A+1D): both images and both statuses coincide with primary; disputed any-progress adds only the already-observed inspect\_haiku45 > original\_sonnet4 outcome. Minimal two-specification witness: mean-checkpoint-progress: inspect\_haiku45; strict-task-completion: original\_sonnet4.

\paragraph{Primary-family pairwise stability.} 0/1 method pairs are stable; the review envelope also has 0/1 stable pairs.
\begin{center}\scriptsize
\begin{tabularx}{\linewidth}{@{}p{0.34\linewidth}Xp{0.09\linewidth}@{}}
\toprule Pair & Relations by specification & Stable? \\
\midrule
inspect\_haiku45 vs. original\_sonnet4 & <: strict-task-completion; >: mean-checkpoint-progress & False \\
\bottomrule\end{tabularx}\end{center}

\paragraph{Residual missing material.} Raw model outputs, task services, and judge traces were not published in this frozen source package. Container base images are tag-bound rather than digest-bound in several task files.

\subsection{BBEH: complete finite-family audit}\label{app:r8-bbeh}

\paragraph{Typed record.} Query: Which published GPT-4o evaluation artifact is higher over the same 23 tasks, and is the order aggregation-stable? Endpoint: ordinal ranking of two fixed per-task accuracy vectors. Fixed substrate (frozen\_summary\_vectors\_plus\_exact\_denominators): 23 table rows plus the pinned 4,520-example dataset; every rounded score implies a unique integer correct count. The audit evaluates every admitted member of the declared finite family; it does not certify family completeness.

\begingroup
\scriptsize
\begin{tabularx}{\linewidth}{@{}p{0.27\linewidth}p{0.14\linewidth}p{0.18\linewidth}X@{}}
\toprule
Specification & Admission & Origin & Semantic coordinate \\
\midrule
harmonic-plus-0-01 & admitted & source-mandated & aggregation and zero-handling offset \\
task-macro-arithmetic & admitted & audit-proposed & aggregation across tasks \\
sample-micro & admitted & source-documented & sample versus task weighting \\
harmonic-no-pseudocount & disputed & audit-proposed & empty/zero case in harmonic aggregation \\
\bottomrule
\end{tabularx}
\endgroup

\paragraph{Exact outputs.} No additional rounding of the reconstructed values is used to determine the claims.

\begingroup
\scriptsize
\begin{longtable}{@{}p{0.25\linewidth}p{0.39\linewidth}p{0.27\linewidth}@{}}
\toprule
System/artifact & Specification output & Exact reconstructed value \\
\midrule
\endfirsthead
\toprule System/artifact & Specification output & Exact reconstructed value \\
\midrule
\endhead
inspect\_reproduction & harmonic\_no\_pseudocount & \texttt{0.0425272071389958} \\
inspect\_reproduction & harmonic\_plus\_0\_01 & \texttt{0.07898405712442381} \\
inspect\_reproduction & sample\_micro & \texttt{0.2150442477876106} \\
inspect\_reproduction & task\_macro\_arithmetic & \texttt{0.22086956521739132} \\
original\_paper & harmonic\_no\_pseudocount & \texttt{0.0} \\
original\_paper & harmonic\_plus\_0\_01 & \texttt{0.060462587043807435} \\
original\_paper & sample\_micro & \texttt{0.22300884955752212} \\
original\_paper & task\_macro\_arithmetic & \texttt{0.22811594202898552} \\
\bottomrule
\end{longtable}
\endgroup

\paragraph{Primary and review-envelope identified sets.} Primary (3A): winner and complete order are non-identified; the winner image is \{inspect\_reproduction\}; \{original\_paper\}, and the order image is inspect\_reproduction > original\_paper; original\_paper > inspect\_reproduction. Review envelope (3A+1D): both images and both statuses coincide with primary; disputed harmonic-no-pseudocount adds only the already-observed inspect\_reproduction > original\_paper outcome. Minimal two-specification witness: harmonic-plus-0-01: inspect\_reproduction; task-macro-arithmetic: original\_paper.

\paragraph{Primary-family pairwise stability.} 0/1 method pairs are stable; the review envelope also has 0/1 stable pairs.
\begin{center}\scriptsize
\begin{tabularx}{\linewidth}{@{}p{0.34\linewidth}Xp{0.09\linewidth}@{}}
\toprule Pair & Relations by specification & Stable? \\
\midrule
inspect\_reproduction vs. original\_paper & <: task-macro-arithmetic, sample-micro; >: harmonic-plus-0-01 & False \\
\bottomrule\end{tabularx}\end{center}

\clearpage
\subsection{Personality: complete finite-family audit}\label{app:r8-personality}

\paragraph{Typed record.} Query: Which reported reproduction profile is closest to its paper reference, and is the order fidelity-definition stable? Endpoint: ordinal ranking of three fixed observed/reference eight-trait profile pairs. Fixed substrate (frozen\_summary\_vectors): Three 8-trait tables for a common 1,600-item seed-41 TRAIT slice. The audit evaluates every admitted member of the declared finite family; it does not certify family completeness.

\begingroup
\scriptsize
\begin{tabularx}{\linewidth}{@{}p{0.27\linewidth}p{0.14\linewidth}p{0.18\linewidth}X@{}}
\toprule
Specification & Admission & Origin & Semantic coordinate \\
\midrule
mae-all-eight & admitted & source-documented & aggregation loss \\
rmse-all-eight & admitted & audit-proposed & aggregation loss \\
pearson-all-eight & admitted & source-documented & absolute level versus profile shape \\
rank-agreement-all-eight & admitted & source-documented & absolute level versus ordinal profile shape \\
mae-big-five-only & disputed & audit-proposed & support eligibility \\
\bottomrule
\end{tabularx}
\endgroup

\paragraph{Exact outputs.} No additional rounding of the reconstructed values is used to determine the claims.

\begingroup
\scriptsize
\begin{longtable}{@{}p{0.25\linewidth}p{0.39\linewidth}p{0.27\linewidth}@{}}
\toprule
System/artifact & Specification output & Exact reconstructed value \\
\midrule
\endfirsthead
\toprule System/artifact & Specification output & Exact reconstructed value \\
\midrule
\endhead
claude3\_opus & negative\_mae\_all8 & \texttt{-4.825000000000001} \\
claude3\_opus & negative\_mae\_big5 & \texttt{-7.300000000000002} \\
claude3\_opus & negative\_rmse\_all8 & \texttt{-6.252199612936236} \\
claude3\_opus & pairwise\_rank\_agreement\_all8 & \texttt{1.0} \\
claude3\_opus & pearson\_all8 & \texttt{0.9977220344447256} \\
gpt35\_turbo & negative\_mae\_all8 & \texttt{-4.04125} \\
gpt35\_turbo & negative\_mae\_big5 & \texttt{-4.4} \\
gpt35\_turbo & negative\_rmse\_all8 & \texttt{-5.5196569186861595} \\
gpt35\_turbo & pairwise\_rank\_agreement\_all8 & \texttt{1.0} \\
gpt35\_turbo & pearson\_all8 & \texttt{0.9874454953821219} \\
gpt4\_turbo & negative\_mae\_all8 & \texttt{-5.53625} \\
gpt4\_turbo & negative\_mae\_big5 & \texttt{-5.74} \\
gpt4\_turbo & negative\_rmse\_all8 & \texttt{-6.357142636436594} \\
gpt4\_turbo & pairwise\_rank\_agreement\_all8 & \texttt{1.0} \\
gpt4\_turbo & pearson\_all8 & \texttt{0.9902645463151414} \\
\bottomrule
\end{longtable}
\endgroup

\paragraph{Primary and review-envelope identified sets.} Primary (4A): winner and complete order are non-identified; the winner image is \{gpt35\_turbo\}; \{claude3\_opus\}; \{claude3\_opus, gpt35\_turbo, gpt4\_turbo\}, and the order image is gpt35\_turbo > claude3\_opus > gpt4\_turbo; claude3\_opus > gpt4\_turbo > gpt35\_turbo; claude3\_opus = gpt35\_turbo = gpt4\_turbo. Review envelope (4A+1D): the winner image and both statuses are unchanged; disputed mae-big-five-only adds the order gpt35\_turbo > gpt4\_turbo > claude3\_opus and hence a new GPT4-over-Claude relation. Minimal two-specification witness: mae-all-eight: gpt35\_turbo; pearson-all-eight: claude3\_opus.

\paragraph{Primary-family pairwise stability.} 0/3 method pairs are stable; the review envelope also has 0/3 stable pairs.
\begin{center}\scriptsize
\begin{tabularx}{\linewidth}{@{}p{0.34\linewidth}Xp{0.09\linewidth}@{}}
\toprule Pair & Relations by specification & Stable? \\
\midrule
claude3\_opus vs. gpt35\_turbo & <: mae-all-eight, rmse-all-eight; =: rank-agreement-all-eight; >: pearson-all-eight & False \\
claude3\_opus vs. gpt4\_turbo & =: rank-agreement-all-eight; >: mae-all-eight, rmse-all-eight, pearson-all-eight & False \\
gpt35\_turbo vs. gpt4\_turbo & <: pearson-all-eight; =: rank-agreement-all-eight; >: mae-all-eight, rmse-all-eight & False \\
\bottomrule\end{tabularx}\end{center}

\paragraph{Residual missing material.} Raw 1,600-item outputs are absent and the pinned TRAIT dataset is auto-gated. The dataset card declares no license in captured metadata. Correlation from rounded printed profiles can differ from correlation computed on underlying unrounded values.

\section{Frozen Random Ten: Seven Zero-Replacement Stopping Records}\label{app:r8-random-stops}

All seven drawn objects remain in the census.  Each record exposes the proposed family and typed endpoint, the precise stopping layer, observed and missing material, the no-imputation rationale, and the minimum unblocking artifact.  These are positive infrastructure findings: replacement would condition the sample on auditability and erase the evidence burden needed for future benchmark release practice.

\subsection{GDM InterCode CTF: protocol stopping record}\label{app:r8-gdm-intercode-ctf}

\paragraph{Typed record.} Query: Which documented attempted system has the highest 78-task flag-recovery rate, with DNF eligibility stated explicitly? Candidate endpoint: ordinal ordering of attempted systems by exact-flag task pass rate. Fixed substrate (incomplete): README reports two complete aggregate scores and one DNF, but no frozen per-task eval logs for those runs.

\begingroup
\scriptsize
\begin{tabularx}{\linewidth}{@{}p{0.27\linewidth}p{0.14\linewidth}p{0.18\linewidth}X@{}}
\toprule
Specification & Admission & Origin & Semantic coordinate \\
\midrule
inspect-78-task-micro & admitted & source mandated & eligibility and denominator \\
all-attempted-dnf-unknown & disputed & explicitly documented alternative & eligibility \\
dnf-as-zero & excluded & disputed & empty-case and eligibility \\
\bottomrule
\end{tabularx}
\endgroup

\paragraph{Stopping layer.} Last completed stage: semantic candidates. Primary mode: comparative observation absent. Secondary modes: nonignorable missing outcome.

\paragraph{Observed public material.} Two completed aggregate accuracies, one Gemini DNF after 42 completed samples, task/evaluator source, and pinned InterCode data binding.

\paragraph{Missing material required for an identified set.}
\begin{itemize}
  \item per-task pass/fail vectors for the two completed runs
  \item Gemini per-task results and a declared handling rule for the interrupted remainder
\end{itemize}

\paragraph{Why no imputation.} DNF is neither zero nor a completed 78-task accuracy; excluding it changes eligibility while zero-filling invents failures.

\paragraph{Why a fresh run is not reconstruction.} Current models, react solver behavior, prompt nudge, and tool/runtime state would produce a new agent-evaluation substrate.

\paragraph{Minimum unblock.} A frozen per-task result table or Inspect eval logs for each attempted run, including explicit missingness for unfinished tasks.

\paragraph{Scientific value.} Exposes eligibility and DNF handling as an identification boundary rather than rewarding only benchmarks that publish complete logs.

\subsection{FrontierScience: protocol stopping record}\label{app:r8-frontierscience}

\paragraph{Typed record.} Query: Which system has the highest expert-science performance when Olympic and Research formats and judge choices are declared? Candidate endpoint: format-stratified score vector or an explicitly justified mixed-format scalar. Fixed substrate (absent\_comparative\_outputs): Source includes a one-model 15-sample-per-format demonstration, not a comparative multi-system output substrate.

\begingroup
\scriptsize
\begin{tabularx}{\linewidth}{@{}p{0.27\linewidth}p{0.14\linewidth}p{0.18\linewidth}X@{}}
\toprule
Specification & Admission & Origin & Semantic coordinate \\
\midrule
format-stratified-vector & admitted & source mandated & format and reference equivalence \\
mixed-format-micro & disputed & explicitly documented alternative & aggregation across formats \\
subject-macro & disputed & defensible extension & aggregation \\
alternate-fixed-grader & admitted conditionally & explicitly documented alternative & reference equivalence \\
\bottomrule
\end{tabularx}
\endgroup

\paragraph{Stopping layer.} Last completed stage: semantic candidates. Primary mode: model judge trace absent. Secondary modes: comparative observation absent; endpoint target fragmentation.

\paragraph{Observed public material.} Full scorer prompts/code, pinned 160-example dataset metadata, and a one-model 15-sample-per-format demonstration with three epochs.

\paragraph{Missing material required for an identified set.}
\begin{itemize}
  \item fixed completions from two or more compared systems
  \item per-sample judge outputs with exact grader model and configuration
  \item a controlling mixed-format versus stratified claim query
\end{itemize}

\paragraph{Why no imputation.} Olympic binary judgments and Research normalized rubric scores are not exchangeable missing values, and judge outputs cannot be inferred from aggregate means.

\paragraph{Why a fresh run is not reconstruction.} A new grader call adds provider/model-version and stochastic judge observations; the demonstration is not the full 160-example task.

\paragraph{Minimum unblock.} Per-sample completions and grader decisions for a common system set, with format, subject, epoch, grader identity, and prompt/config bindings.

\paragraph{Scientific value.} Shows that public evaluator code is insufficient when reference equivalence is itself a model-mediated observation.

\subsection{Agent Bench: protocol stopping record}\label{app:r8-agent-bench}

\paragraph{Typed record.} Query: Which agent has the highest OS-task success over the frozen 26-task test split? Candidate endpoint: task-micro binary sandbox-check accuracy. Fixed substrate (absent\_comparative\_outputs): Dataset and executable checks are frozen, but no public multi-agent output/sandbox-state logs are bundled.

\begingroup
\scriptsize
\begin{tabularx}{\linewidth}{@{}p{0.27\linewidth}p{0.14\linewidth}p{0.18\linewidth}X@{}}
\toprule
Specification & Admission & Origin & Semantic coordinate \\
\midrule
binary-task-micro & admitted & source mandated & aggregation \\
checker-family-vector & admitted & defensible extension & checker type \\
dev-plus-test & excluded & disputed & eligibility \\
\bottomrule
\end{tabularx}
\endgroup

\paragraph{Stopping layer.} Last completed stage: semantic candidates. Primary mode: stateful execution artifact absent. Secondary modes: comparative observation absent.

\paragraph{Observed public material.} Embedded 26-task test data, pinned AgentBench commit, digest-pinned base image, and executable task-specific check logic.

\paragraph{Missing material required for an identified set.}
\begin{itemize}
  \item final sandbox filesystem/process state for each evaluated agent
  \item submitted answers and task-level checker outputs for a common system set
\end{itemize}

\paragraph{Why no imputation.} The evaluator executes scripts against final state; a text answer alone does not determine task success.

\paragraph{Why a fresh run is not reconstruction.} Re-running agents recreates mutable sandbox trajectories and model observations rather than evaluating preserved states.

\paragraph{Minimum unblock.} Restorable per-task sandbox snapshots or a sufficient state manifest plus checker stdout, return code, submission, and exact image digest.

\paragraph{Scientific value.} Identifies state preservation, not merely source release, as the reproducibility unit for execution-based agent benchmarks.

\subsection{CORE-Bench: protocol stopping record}\label{app:r8-core-bench}

\paragraph{Typed record.} Query: Which agent fully answers the frozen scientific-reproduction task questions under a declared difficulty and eligibility rule? Candidate endpoint: capsule-task binary all-question success, optionally paired with question-level partial credit. Fixed substrate (absent\_comparative\_outputs): Source and encrypted metadata are public, but final report.json files and agent trajectories are absent.

\begingroup
\scriptsize
\begin{tabularx}{\linewidth}{@{}p{0.27\linewidth}p{0.14\linewidth}p{0.18\linewidth}X@{}}
\toprule
Specification & Admission & Origin & Semantic coordinate \\
\midrule
all-question-task-success & admitted & source mandated & partial credit \\
question-micro-partial & disputed & defensible extension & partial credit and denominator \\
capsule-macro-partial & disputed & defensible extension & aggregation \\
easy-only-versus-all-difficulties & excluded from same family & explicitly documented alternative & eligibility \\
\bottomrule
\end{tabularx}
\endgroup

\paragraph{Stopping layer.} Last completed stage: semantic candidates. Primary mode: stateful execution artifact absent. Secondary modes: comparative observation absent; resource envelope mismatch.

\paragraph{Observed public material.} Scorer and capsule bindings, encrypted public dataset metadata, per-capsule checksums, and declared easy/medium/hard task construction.

\paragraph{Missing material required for an identified set.}
\begin{itemize}
  \item final report.json outputs for two or more agents
  \item task/capsule eligibility ledger after GPU, vision, and download filters
  \item question-level results needed for partial-credit alternatives
\end{itemize}

\paragraph{Why no imputation.} Binary all-question success and question-level partial credit cannot be recovered from absent report files or aggregate accuracy.

\paragraph{Why a fresh run is not reconstruction.} Full replay needs about 15 GB of capsules plus Docker overhead; rerunning agents would create new observations and a new substrate rather than reconstruct the frozen output.

\paragraph{Minimum unblock.} Frozen report.json files, exact eligible capsule/question manifest, and scorer explanations for each system; sandbox images only if state re-evaluation is requested.

\paragraph{Scientific value.} Separates an output-preservation failure from a compute failure and quantifies when resource limits would bias a convenience-only audit.

\subsection{MMIU: protocol stopping record}\label{app:r8-mmiu}

\paragraph{Typed record.} Query: Which vision-language system is best over the frozen 11,698-example, 52-task MMIU test set under declared weighting? Candidate endpoint: ordinal ranking by overall or task-balanced multiple-choice accuracy. Fixed substrate (absent\_comparative\_outputs): Pinned dataset is public but about 25.5 GB; no comparative response logs are bundled.

\begingroup
\scriptsize
\begin{tabularx}{\linewidth}{@{}p{0.27\linewidth}p{0.14\linewidth}p{0.18\linewidth}X@{}}
\toprule
Specification & Admission & Origin & Semantic coordinate \\
\midrule
example-micro-accuracy & admitted & source mandated & aggregation \\
task-macro-accuracy & admitted & explicitly documented alternative & aggregation \\
relationship-type-macro & disputed & defensible extension & aggregation \\
\bottomrule
\end{tabularx}
\endgroup

\paragraph{Stopping layer.} Last completed stage: semantic candidates. Primary mode: comparative observation absent. Secondary modes: resource envelope mismatch.

\paragraph{Observed public material.} Pinned dataset/model-input bindings, exact ZIP checksums, choice scorer, and task/category aggregation code.

\paragraph{Missing material required for an identified set.}
\begin{itemize}
  \item fixed answer choices for two or more vision-language systems
  \item per-example task/category metadata joined to those answers
\end{itemize}

\paragraph{Why no imputation.} Aggregate leaderboard values cannot be disaggregated into example-micro, task-macro, or relationship-type macro scores.

\paragraph{Why a fresh run is not reconstruction.} The pinned HF repository is about 25.5 GB, and fresh VLM inference would create new observations rather than reconstruct the frozen comparative substrate.

\paragraph{Minimum unblock.} Per-example answer files with sample IDs and model/version/prompt bindings; images need not be redistributed if answer keys and metadata joins are independently verifiable.

\paragraph{Scientific value.} Demonstrates that large public datasets can remain unauditable at the claim layer when compact response artifacts are not released.

\subsection{MMMU: protocol stopping record}\label{app:r8-mmmu}

\paragraph{Typed record.} Query: Which multimodal system is best when multiple-choice and open-ended MMMU formats and their evaluator paths are declared? Candidate endpoint: format-stratified accuracy vector; mixed scalar only with an explicit format aggregation. Fixed substrate (absent\_comparative\_outputs): Pinned 3.36 GB dataset is public; no fixed system completions or model-judge outputs are bundled.

\begingroup
\scriptsize
\begin{tabularx}{\linewidth}{@{}p{0.27\linewidth}p{0.14\linewidth}p{0.18\linewidth}X@{}}
\toprule
Specification & Admission & Origin & Semantic coordinate \\
\midrule
format-stratified-vector & admitted & source mandated & reference equivalence and format \\
combined-example-micro & disputed & explicitly documented alternative & aggregation and denominator \\
equal-format-macro & disputed & defensible extension & aggregation \\
\bottomrule
\end{tabularx}
\endgroup

\paragraph{Stopping layer.} Last completed stage: semantic candidates. Primary mode: model judge trace absent. Secondary modes: comparative observation absent; endpoint target fragmentation.

\paragraph{Observed public material.} Pinned 847 multiple-choice and 53 open examples, choice evaluator, model-graded open evaluator, and format-specific task definitions.

\paragraph{Missing material required for an identified set.}
\begin{itemize}
  \item fixed system responses for both formats
  \item open-answer judge outputs and exact judge binding
  \item a declared rule for micro versus equal-format aggregation
\end{itemize}

\paragraph{Why no imputation.} The 53 open items cannot be scored from absent model-judge traces, and separate format accuracies do not determine a unique combined endpoint without weights.

\paragraph{Why a fresh run is not reconstruction.} New model-judge calls create additional semantic/stochastic observations; new system completions are not the published substrate.

\paragraph{Minimum unblock.} Sample-keyed completions and scores for both formats, including judge prompt/model/config/output for every open item.

\paragraph{Scientific value.} Makes evaluator-mediated reference equivalence and cross-format aggregation jointly visible instead of collapsing both into an unavailable scalar.

\subsection{BOLD: protocol stopping record}\label{app:r8-bold}

\paragraph{Typed record.} Query: How does one fixed generator's BOLD behavior vary by domain under toxicity, sentiment, and regard endpoints without treating them as interchangeable? Candidate endpoint: three separately typed domain-score vectors. Fixed substrate (one\_reported\_system\_summary\_only): README reports one GPT-4o-mini domain table but no raw 7,200 completions.

\begingroup
\scriptsize
\begin{tabularx}{\linewidth}{@{}p{0.27\linewidth}p{0.14\linewidth}p{0.18\linewidth}X@{}}
\toprule
Specification & Admission & Origin & Semantic coordinate \\
\midrule
toxicity-any-class-0-3 & admitted & source mandated & threshold and reference model \\
sentiment-three-way-0-5 & separate endpoint & source mandated & threshold \\
regard-class-score & separate endpoint & source mandated & reference model and label mapping \\
original-unreleased-toxicity-classifier & not executable & explicitly documented alternative & reference equivalence \\
\bottomrule
\end{tabularx}
\endgroup

\paragraph{Stopping layer.} Last completed stage: semantic candidates. Primary mode: endpoint target fragmentation. Secondary modes: comparative observation absent; evaluator reference unavailable.

\paragraph{Observed public material.} One GPT-4o-mini domain summary for toxicity, sentiment, and regard; pinned current dataset and regard-model metadata; current scorer code.

\paragraph{Missing material required for an identified set.}
\begin{itemize}
  \item raw 7,200 generated completions
  \item outputs for a second compared generator for a winner/order claim
  \item original paper's unreleased toxicity classifier outputs
\end{itemize}

\paragraph{Why no imputation.} Toxicity, sentiment, and regard measure different constructs; their aggregate domain scores cannot be treated as alternative values of one common endpoint.

\paragraph{Why a fresh run is not reconstruction.} Current Detoxify/regard replay on newly generated text would not recover the printed GPT-4o-mini outputs or the unreleased original classifier.

\paragraph{Minimum unblock.} Frozen sample-keyed completions for at least two systems plus raw current-classifier outputs; original-classifier results must remain unavailable unless authors release them.

\paragraph{Scientific value.} Produces a positive boundary result: not every documented metric set is a semantic family, and forcing one would create a false identified-set exercise.  The contract is therefore not monotone toward admission: when candidates target distinct constructs rather than alternative realizations of one typed query, it stops instead of manufacturing a family.

\section{Three Purposive Deep Audits}\label{app:r8-deep-three}

The selection mechanism limits only probability-denominator inclusion.  It does not limit the analytical importance of three fully enumerated, domain-distinct audits.  Under their declared families all three preserve the winner across both admission scopes, but their order results differ: AgentDojo and AutoML are identified in the primary family and non-identified in the review envelope, whereas BEIR SciFact is non-identified in both.  A separate BEIR evaluator-grid sensitivity deliberately tests, and rejects, transport of the five-view winner and backbone to a wider family.  This is a purposive cross-domain mechanism comparison, not prevalence, independent replication, or a family-completeness certificate.  The headline query is ordinal, so displayed per-specification values are not pooled into a cross-specification exact-score identified set.  In the worked tables, SM denotes source-mandated origin, SE source/evaluator-exposed, and AP audit-proposed; A enters the primary family and D enters the admitted-plus-disputed review envelope only.  Excluded candidates enter neither; no code certifies family completeness.

\Needspace{26\baselineskip}
\subsection{AgentDojo: purposive deep audit}\label{app:r8-agentdojo-v0-1-35}

\paragraph{Typed record.} Frozen union of 949 public important\_instructions run keys for three models; 629 keys are shared. Predictions/trajectories and boolean utility/security outputs are fixed; only eligibility and aggregation change. Endpoint type: ranking; higher is better: False; tie tolerance: 1e-10. The purposive selection mechanism excludes this object from the random-sample probability denominator, but does not reduce the analytical importance of the fully enumerated mechanism audit.

\begingroup\scriptsize
\setlength{\tabcolsep}{2pt}
\begin{tabularx}{\linewidth}{@{}p{0.22\linewidth}p{0.055\linewidth}p{0.055\linewidth}p{0.20\linewidth}X@{}}
\toprule Specification & Origin & Adm. & Coordinate & Exact outputs by system \\
\midrule
published\_v1\_1\_2\_micro\_asr & SM & A & Published v1.1.2 micro targeted ASR & claude-3-5-sonnet-20241022=1.1128775835; gemini-2.0-flash-001=20.826709062; gpt-4o-mini-2024-07-18=27.186009539 \\
all\_available\_run\_micro\_asr & AP & A & All-available-run micro targeted ASR & claude-3-5-sonnet-20241022=1.1128775835; gemini-2.0-flash-001=14.1201264489; gpt-4o-mini-2024-07-18=27.186009539 \\
common\_key\_suite\_macro\_asr & AP & D & Common-key suite-macro targeted ASR & claude-3-5-sonnet-20241022=1.4732142857; gemini-2.0-flash-001=26.7311507936; gpt-4o-mini-2024-07-18=30.6646825397 \\
utility\_conditional\_asr & AP & D & ASR conditional on user-task utility success & claude-3-5-sonnet-20241022=1.3157894737; gemini-2.0-flash-001=32; gpt-4o-mini-2024-07-18=19.7452229299 \\
worst\_suite\_asr & AP & D & Worst-suite common-key targeted ASR & claude-3-5-sonnet-20241022=3.8095238095; gemini-2.0-flash-001=66.6666666667; gpt-4o-mini-2024-07-18=57.1428571429 \\
\bottomrule\end{tabularx}\endgroup

\paragraph{Displayed specification values.} The public v0.3 endpoint is ordinal; these values remain auditable inputs but are not pooled into a cross-specification exact-score identified set.
\begingroup\scriptsize
\begin{tabularx}{\linewidth}{@{}p{0.28\linewidth}X@{}}
\toprule System & Values by specification \\
\midrule
claude-3-5-sonnet-20241022 & 1.1128775835 [published\_v1\_1\_2\_micro\_asr, all\_available\_run\_micro\_asr]; 1.3157894737 [utility\_conditional\_asr]; 1.4732142857 [common\_key\_suite\_macro\_asr]; 3.8095238095 [worst\_suite\_asr] \\
gemini-2.0-flash-001 & 14.1201264489 [all\_available\_run\_micro\_asr]; 20.826709062 [published\_v1\_1\_2\_micro\_asr]; 26.7311507936 [common\_key\_suite\_macro\_asr]; 32 [utility\_conditional\_asr]; 66.6666666667 [worst\_suite\_asr] \\
gpt-4o-mini-2024-07-18 & 19.7452229299 [utility\_conditional\_asr]; 27.186009539 [published\_v1\_1\_2\_micro\_asr, all\_available\_run\_micro\_asr]; 30.6646825397 [common\_key\_suite\_macro\_asr]; 57.1428571429 [worst\_suite\_asr] \\
\bottomrule\end{tabularx}\endgroup

\paragraph{Primary and review-envelope claims.} Primary (2A): winner and complete order are identified, with order claude-3-5-sonnet-20241022  >  gemini-2.0-flash-001  >  gpt-4o-mini-2024-07-18.  Review envelope (2A+3D): the winner remains identified as \{claude-3-5-sonnet-20241022\}, while the complete order is non-identified and additionally contains claude-3-5-sonnet-20241022  >  gpt-4o-mini-2024-07-18  >  gemini-2.0-flash-001.  Published\_v1\_1\_2\_micro\_asr versus utility\_conditional\_asr is a review-envelope witness.

\Needspace{20\baselineskip}
\paragraph{Pairwise stability.} Primary: 3/3 pairs stable over 2A.  Review envelope: 2/3 stable over 2A+3D.  The matrix below reports the envelope.
\begingroup\scriptsize
\begin{tabularx}{\linewidth}{@{}p{0.34\linewidth}Xp{0.09\linewidth}@{}}
\toprule Pair & Relations by specification & Stable? \\
\midrule
claude-3-5-sonnet-20241022 vs. gemini-2.0-flash-001 & >: published\_v1\_1\_2\_micro\_asr, all\_available\_run\_micro\_asr, common\_key\_suite\_macro\_asr, utility\_conditional\_asr, worst\_suite\_asr & True \\
claude-3-5-sonnet-20241022 vs. gpt-4o-mini-2024-07-18 & >: published\_v1\_1\_2\_micro\_asr, all\_available\_run\_micro\_asr, common\_key\_suite\_macro\_asr, utility\_conditional\_asr, worst\_suite\_asr & True \\
gemini-2.0-flash-001 vs. gpt-4o-mini-2024-07-18 & <: utility\_conditional\_asr, worst\_suite\_asr; >: published\_v1\_1\_2\_micro\_asr, all\_available\_run\_micro\_asr, common\_key\_suite\_macro\_asr & False \\
\bottomrule\end{tabularx}\endgroup

\Needspace{26\baselineskip}
\subsection{AutoML Benchmark: purposive deep audit}\label{app:r8-automlbenchmark-results-valid}

\paragraph{Typed record.} Frozen 149-row public results\_valid.csv with 5 frameworks, 3 binary-AUC tasks, and one missing framework--task--fold cell; AUC values and the missingness mask are fixed. Endpoint type: ranking; higher is better: True; tie tolerance: 1e-10. The purposive selection mechanism excludes this object from the random-sample probability denominator, but does not reduce the analytical importance of the fully enumerated mechanism audit.

\begingroup\scriptsize
\setlength{\tabcolsep}{2pt}
\begin{tabularx}{\linewidth}{@{}p{0.22\linewidth}p{0.055\linewidth}p{0.055\linewidth}p{0.20\linewidth}X@{}}
\toprule Specification & Origin & Adm. & Coordinate & Exact outputs by system \\
\midrule
available\_fold\_micro\_auc & AP & A & Available-fold micro mean AUC & AutoWEKA=0.7803311667; H2OAutoML=0.8357915667; TPOT=0.8039506; autosklearn=0.8297094828; oboe=0.7597841667 \\
task\_macro\_auc & AP & A & Task-macro mean AUC & AutoWEKA=0.7803311667; H2OAutoML=0.8357915667; TPOT=0.8039506; autosklearn=0.8224988889; oboe=0.7597841667 \\
\path{complete_case_fold_micro_auc} & AP & A & Complete-case common-fold micro mean AUC & AutoWEKA=0.7864872069; H2OAutoML=0.8412271034; TPOT=0.8141202414; autosklearn=0.8297094828; oboe=0.7672134483 \\
negative\_mean\_task\_rank & SM & A & Source-prescribed average task rank & AutoWEKA=-4; H2OAutoML=-1.3333333333; TPOT=-2.6666666667; autosklearn=-2.3333333333; oboe=-4.6666666667 \\
worst\_task\_auc & AP & D & Worst-task mean AUC & AutoWEKA=0.599918; H2OAutoML=0.6407225; TPOT=0.557225; autosklearn=0.6133916667; oboe=0.5397373 \\
\bottomrule\end{tabularx}\endgroup

\paragraph{Canonical procedure and subset realization.} AMLB prescribes replacing missing values by constant-predictor performance, averaging folds within each task, ranking frameworks within task, and comparing average ranks across tasks \citep[Sec.~6.1]{GijsbersEtAl2024AMLB}.  On this frozen typed subset the sole missing cell is \texttt{dresses-sales/autosklearn/fold-5}; binary-AUC constant-predictor performance is 0.5.  Recomputing all task means and ranks yields exactly the displayed negative average ranks.  The realization uses five frameworks and three legacy binary-AUC tasks rather than AMLB's full 2024 study and imports neither its Friedman nor Nemenyi inference.  Correcting the candidate's origin/admission therefore changes provenance, not the identified winner, order, or pair relations.

\paragraph{Displayed specification values.} The public v0.3 endpoint is ordinal; these values remain auditable inputs but are not pooled into a cross-specification exact-score identified set.
\begingroup\scriptsize
\begin{tabularx}{\linewidth}{@{}p{0.28\linewidth}X@{}}
\toprule System & Values by specification \\
\midrule
AutoWEKA & -4 [negative\_mean\_task\_rank]; 0.599918 [worst\_task\_auc]; 0.7803311667 [available\_fold\_micro\_auc, task\_macro\_auc]; 0.7864872069 [complete\_case\_fold\_micro\_auc] \\
H2OAutoML & -1.3333333333 [negative\_mean\_task\_rank]; 0.6407225 [worst\_task\_auc]; 0.8357915667 [available\_fold\_micro\_auc, task\_macro\_auc]; 0.8412271034 [complete\_case\_fold\_micro\_auc] \\
TPOT & -2.6666666667 [negative\_mean\_task\_rank]; 0.557225 [worst\_task\_auc]; 0.8039506 [available\_fold\_micro\_auc, task\_macro\_auc]; 0.8141202414 [complete\_case\_fold\_micro\_auc] \\
autosklearn & -2.3333333333 [negative\_mean\_task\_rank]; 0.6133916667 [worst\_task\_auc]; 0.8224988889 [task\_macro\_auc]; 0.8297094828 [available\_fold\_micro\_auc, complete\_case\_fold\_micro\_auc] \\
oboe & -4.6666666667 [negative\_mean\_task\_rank]; 0.5397373 [worst\_task\_auc]; 0.7597841667 [available\_fold\_micro\_auc, task\_macro\_auc]; 0.7672134483 [complete\_case\_fold\_micro\_auc] \\
\bottomrule\end{tabularx}\endgroup

\paragraph{Primary and review-envelope claims.} Primary (4A): winner and complete order are identified, with order H2OAutoML  >  autosklearn  >  TPOT  >  AutoWEKA  >  oboe.  Review envelope (4A+1D): the winner remains identified as \{H2OAutoML\}, while the complete order is non-identified and additionally contains H2OAutoML  >  autosklearn  >  AutoWEKA  >  TPOT  >  oboe.  Available\_fold\_micro\_auc versus worst\_task\_auc is a review-envelope witness.

\Needspace{32\baselineskip}
\paragraph{Pairwise stability.} Primary: 10/10 pairs stable over 4A.  Review envelope: 9/10 stable over 4A+1D.  The matrix below reports the envelope.
\begingroup\scriptsize
\begin{tabularx}{\linewidth}{@{}p{0.34\linewidth}Xp{0.09\linewidth}@{}}
\toprule Pair & Relations by specification & Stable? \\
\midrule
AutoWEKA vs. H2OAutoML & <: available\_fold\_micro\_auc, task\_macro\_auc, complete\_case\_fold\_micro\_auc, negative\_mean\_task\_rank, worst\_task\_auc & True \\
AutoWEKA vs. TPOT & <: available\_fold\_micro\_auc, task\_macro\_auc, complete\_case\_fold\_micro\_auc, negative\_mean\_task\_rank; >: worst\_task\_auc & False \\
AutoWEKA vs. autosklearn & <: available\_fold\_micro\_auc, task\_macro\_auc, complete\_case\_fold\_micro\_auc, negative\_mean\_task\_rank, worst\_task\_auc & True \\
AutoWEKA vs. oboe & >: available\_fold\_micro\_auc, task\_macro\_auc, complete\_case\_fold\_micro\_auc, negative\_mean\_task\_rank, worst\_task\_auc & True \\
H2OAutoML vs. TPOT & >: available\_fold\_micro\_auc, task\_macro\_auc, complete\_case\_fold\_micro\_auc, negative\_mean\_task\_rank, worst\_task\_auc & True \\
H2OAutoML vs. autosklearn & >: available\_fold\_micro\_auc, task\_macro\_auc, complete\_case\_fold\_micro\_auc, negative\_mean\_task\_rank, worst\_task\_auc & True \\
H2OAutoML vs. oboe & >: available\_fold\_micro\_auc, task\_macro\_auc, complete\_case\_fold\_micro\_auc, negative\_mean\_task\_rank, worst\_task\_auc & True \\
TPOT vs. autosklearn & <: available\_fold\_micro\_auc, task\_macro\_auc, complete\_case\_fold\_micro\_auc, negative\_mean\_task\_rank, worst\_task\_auc & True \\
TPOT vs. oboe & >: available\_fold\_micro\_auc, task\_macro\_auc, complete\_case\_fold\_micro\_auc, negative\_mean\_task\_rank, worst\_task\_auc & True \\
autosklearn vs. oboe & >: available\_fold\_micro\_auc, task\_macro\_auc, complete\_case\_fold\_micro\_auc, negative\_mean\_task\_rank, worst\_task\_auc & True \\
\bottomrule\end{tabularx}\endgroup

\Needspace{26\baselineskip}
\subsection{BEIR SciFact: purposive deep audit}\label{app:r8-beir-scifact}

\paragraph{Typed record.} SciFact test retrieval with 5183 frozen documents, 300 test queries, fixed qrels, three deterministic BM25 runs, and top-100 ranked lists; only the documented metric query changes. Endpoint type: ranking; higher is better: True; tie tolerance: 1e-10. The declared five-view family selects one documented reporting cutoff per metric path: NDCG@10, MAP@100, Recall@100, P@10, and MRR@10.  It is grounded and finite, but not an exhaustive enumeration of evaluator-exposed cutoffs.  The purposive selection mechanism excludes this object from the random-sample probability denominator, but does not reduce the analytical importance of the fully enumerated mechanism audit.

\begingroup\scriptsize
\setlength{\tabcolsep}{2pt}
\begin{tabularx}{\linewidth}{@{}p{0.22\linewidth}p{0.055\linewidth}p{0.055\linewidth}p{0.20\linewidth}X@{}}
\toprule Specification & Origin & Adm. & Coordinate & Exact outputs by system \\
\midrule
ndcg\_at\_10 & SE & A & NDCG@10 & bm25\_k1\_0\_9\_b\_0\_4=0.65883; bm25\_k1\_1\_2\_b\_0\_75=0.66046; bm25\_titlex2\_k1\_1\_2\_b\_0\_75=0.66895 \\
map\_at\_100 & SE & A & MAP@100 & bm25\_k1\_0\_9\_b\_0\_4=0.62252; bm25\_k1\_1\_2\_b\_0\_75=0.61951; bm25\_titlex2\_k1\_1\_2\_b\_0\_75=0.63057 \\
recall\_at\_100 & SE & A & Recall@100 & bm25\_k1\_0\_9\_b\_0\_4=0.88856; bm25\_k1\_1\_2\_b\_0\_75=0.88589; bm25\_titlex2\_k1\_1\_2\_b\_0\_75=0.88922 \\
mrr\_at\_10 & SE & A & MRR@10 & bm25\_k1\_0\_9\_b\_0\_4=0.62877; bm25\_k1\_1\_2\_b\_0\_75=0.62697; bm25\_titlex2\_k1\_1\_2\_b\_0\_75=0.63565 \\
precision\_at\_10 & SE & A & P@10 & bm25\_k1\_0\_9\_b\_0\_4=0.085; bm25\_k1\_1\_2\_b\_0\_75=0.087; bm25\_titlex2\_k1\_1\_2\_b\_0\_75=0.08733 \\
\bottomrule\end{tabularx}\endgroup

\paragraph{Displayed specification values.} The public v0.3 endpoint is ordinal; these values remain auditable inputs but are not pooled into a cross-specification exact-score identified set.
\begingroup\scriptsize
\begin{tabularx}{\linewidth}{@{}p{0.28\linewidth}X@{}}
\toprule System & Values by specification \\
\midrule
bm25\_k1\_0\_9\_b\_0\_4 & 0.085 [precision\_at\_10]; 0.62252 [map\_at\_100]; 0.62877 [mrr\_at\_10]; 0.65883 [ndcg\_at\_10]; 0.88856 [recall\_at\_100] \\
bm25\_k1\_1\_2\_b\_0\_75 & 0.087 [precision\_at\_10]; 0.61951 [map\_at\_100]; 0.62697 [mrr\_at\_10]; 0.66046 [ndcg\_at\_10]; 0.88589 [recall\_at\_100] \\
bm25\_titlex2\_k1\_1\_2\_b\_0\_75 & 0.08733 [precision\_at\_10]; 0.63057 [map\_at\_100]; 0.63565 [mrr\_at\_10]; 0.66895 [ndcg\_at\_10]; 0.88922 [recall\_at\_100] \\
\bottomrule\end{tabularx}\endgroup

\paragraph{Primary and review-envelope claims.} All five source-documented specifications are admitted, so primary and review envelope coincide.  The winner is identified as \{bm25\_titlex2\_k1\_1\_2\_b\_0\_75\}; the complete order is non-identified, with images bm25\_titlex2\_k1\_1\_2\_b\_0\_75  >  bm25\_k1\_1\_2\_b\_0\_75  >  bm25\_k1\_0\_9\_b\_0\_4 and bm25\_titlex2\_k1\_1\_2\_b\_0\_75  >  bm25\_k1\_0\_9\_b\_0\_4  >  bm25\_k1\_1\_2\_b\_0\_75.  ndcg\_at\_10 versus map\_at\_100 is a minimum witness.

\Needspace{20\baselineskip}
\paragraph{Pairwise stability in the declared family.} Primary and review envelope coincide: 2/3 pairs are stable over the five selected views (5A).
\begingroup\scriptsize
\begin{tabularx}{\linewidth}{@{}p{0.34\linewidth}Xp{0.09\linewidth}@{}}
\toprule Pair & Relations by specification & Stable? \\
\midrule
bm25\_k1\_0\_9\_b\_0\_4 vs. bm25\_k1\_1\_2\_b\_0\_75 & <: ndcg\_at\_10, precision\_at\_10; >: map\_at\_100, recall\_at\_100, mrr\_at\_10 & False \\
bm25\_k1\_0\_9\_b\_0\_4 vs. bm25\_titlex2\_k1\_1\_2\_b\_0\_75 & <: ndcg\_at\_10, map\_at\_100, recall\_at\_100, mrr\_at\_10, precision\_at\_10 & True \\
bm25\_k1\_1\_2\_b\_0\_75 vs. bm25\_titlex2\_k1\_1\_2\_b\_0\_75 & <: ndcg\_at\_10, map\_at\_100, recall\_at\_100, mrr\_at\_10, precision\_at\_10 & True \\
\bottomrule\end{tabularx}\endgroup

\paragraph{Evaluator-grid family sensitivity.} The pinned evaluator exposes default cutoffs \(\{1,3,5,10,100,1000\}\).  Because the frozen runs are bound only through depth 100, the executable grid crosses the same five metric paths with \(k\in\{1,3,5,10,100\}\), yielding 25 views; \(k=1000\) is excluded rather than extrapolated.  The five selected cells match the existing official-path cross-check at five decimals.  Over the grid, each of the three systems is a possible winner, five weak orders occur, and every system pair changes relation, so winner and complete order are non-identified and the stable backbone contracts from 2/3 to 0/3.  This sensitivity neither replaces the declared five-view family nor certifies natural family completion.  It is the positive case showing that a backbone is licensed by \((D,\F_{\mathrm{scope}},q)\), not attached to the evaluator in isolation.
\endgroup

\section{Outcome-exposed exploratory landscape}
\label{app:exploratory-eight}

The following cases were selected after outcomes could be inspected.  They map mechanisms and boundaries but do not inherit prospective admission or a shared denominator.

\begingroup\scriptsize
\setlength{\tabcolsep}{2pt}
\begin{longtable}{@{}p{0.12\linewidth}p{0.21\linewidth}p{0.20\linewidth}p{0.21\linewidth}p{0.20\linewidth}@{}}
\caption{Claim-resolution profile for eight exploratory cases.}\label{tab:reviewer-exploratory-hardening}\\
\toprule Case & Compared readings & Candidate status & Claim resolution & Boundary \\
\midrule\endfirsthead
\toprule Case & Compared readings & Candidate status & Claim resolution & Boundary \\
\midrule\endhead
VQA-RAD & overall micro vs equal closed/open macro & second aggregation audit-proposed & exact \NID{}; winner/order \ID{}; 1/1 stable & source does not declare equal answer-type overall \\
MORU & sample-first overall vs equal 16-dimension macro & second aggregation audit-proposed & exact \NID{}; winner \ID{}; order \NID{}; 9/10 stable & dimension vector exposed; second overall not declared \\
InstrumentalEval & prompt-global vs equal six-type macro & second aggregation audit-proposed & single-system exact \NID{} & per-type metrics do not declare a second overall \\
ANIMA & historical overall vs dimension-normalised average & both source-declared & exact \NID{}; winner \ID{}; order \NID{}; 5/6 stable & table-level reconstruction at displayed precision \\
AIR Bench & item-micro vs equal-category macro delta & second aggregation audit-proposed & exact delta \NID{}; direction \ID{} & category rows do not declare a macro headline \\
LiveBench & subtask--category vs item--category macro & both source/evaluator-exposed & exact \NID{}; winner/order \ID{}; 1/1 stable & historical aggregates and displayed precision only \\
UCCB & 1--5 mean vs percentage \(\times20\) & representation-only; domain macro excluded & display changes; winner/order \ID{}; 10/10 stable & reversible unit change is not semantic instability \\
BFCL & 3,399-subset micro vs equal ten-group macro & grouping exposed; overall macro audit-proposed & subset winner/order \NID{}; 0/1 stable & not a 3,981-headline result; support does not close \\
\bottomrule
\end{longtable}
\endgroup

The landscape includes positive boundaries as well as reversals.  UCCB shows why endpoint typing removes a false instability; VQA-RAD and LiveBench retain decisions despite scalar variation; BFCL shows why a support-limited reversal cannot be promoted to the advertised headline.

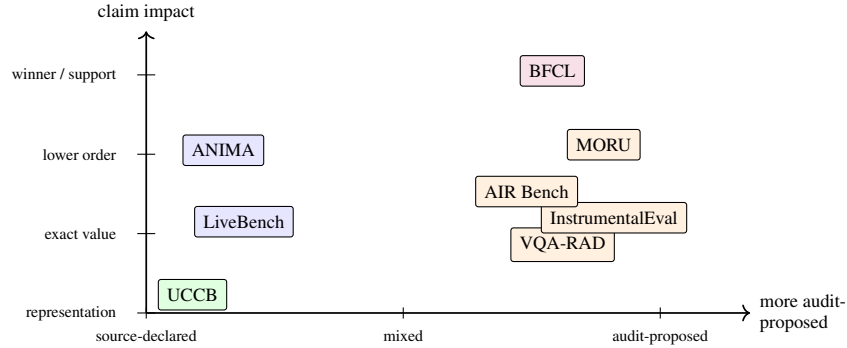
\begin{figure}[ht]
\centering
\begin{tikzpicture}[font=\scriptsize,x=3.4cm,y=1.05cm]
  \draw[->,line width=.65pt] (0,0) -- (2.35,0) node[right,align=left] {more audit-\\proposed};
  \draw[->,line width=.65pt] (0,0) -- (0,3.55) node[above,align=center] {claim impact};
  \foreach \x/\lab in {0/{source-declared},1/{mixed},2/{audit-proposed}} {
    \draw (\x,.07) -- (\x,-.07);
    \node[below,align=center,font=\tiny] at (\x,-.10) {\lab};
  }
  \foreach \y/\lab in {0/{representation},1/{exact value},2/{lower order},3/{winner / support}} {
    \draw (.035,\y) -- (-.035,\y);
    \node[anchor=east,font=\tiny] at (-.08,\y) {\lab};
  }
  \node[draw,rounded corners=1pt,fill=green!12] at (.18,.23) {UCCB};
  \node[draw,rounded corners=1pt,fill=blue!10] at (.38,1.15) {LiveBench};
  \node[draw,rounded corners=1pt,fill=blue!10] at (.30,2.05) {ANIMA};
  \node[draw,rounded corners=1pt,fill=orange!12] at (1.62,.86) {VQA-RAD};
  \node[draw,rounded corners=1pt,fill=orange!12] at (1.82,1.20) {InstrumentalEval};
  \node[draw,rounded corners=1pt,fill=orange!12] at (1.48,1.53) {AIR Bench};
  \node[draw,rounded corners=1pt,fill=orange!12] at (1.78,2.12) {MORU};
  \node[draw,rounded corners=1pt,fill=purple!12] at (1.58,3.05) {BFCL};
\end{tikzpicture}
\caption{\textbf{Mechanism map for the eight exploratory cases.} Position is a qualitative routing summary, not a statistical embedding.  It separates harmless representation changes, exact-value sensitivity, lower-order changes, and the BFCL support-limited winner candidate.}
\label{fig:reviewer-exploratory-map}
\end{figure}

\subsection{Case-level mechanism cards}

\paragraph{VQA-RAD.}
The fixed table reports overall, closed-answer, and open-answer accuracies.  An equal closed/open macro is transparent but not source-declared as an overall.  Exact values differ; the GPT-5.2 winner and the sole pair relation remain fixed.

\paragraph{MORU.}
The source declares a sample-first overall and exposes sixteen dimension means.  Equal dimension weighting is an audit proposal.  It changes only the Gemini/Grok relation, leaving GPT-5.2 the winner and 9/10 pairs stable.

\paragraph{InstrumentalEval.}
The evaluator declares a prompt-global convergence fraction and reports six task-type metrics.  Equal task-type weighting is audit-proposed.  With one bound system this is scalar sensitivity only; winner, order, and pair claims are not applicable.

\paragraph{ANIMA.}
The source explicitly reports a historical overall and a dimension-normalised average.  The latter is reconstructable from thirteen displayed means, while raw prompts, grader replies, and weights are absent.  GPT-5-mini remains the winner; Claude and Gemini swap; 5/6 pairs remain stable.

\paragraph{AIR Bench.}
The fixed source reports a 5,694-item overall delta and sixteen category rows.  Equal category weighting is audit-proposed.  The exact delta changes while its tested-versus-paper direction stays positive.

\paragraph{LiveBench.}
The README exposes subtask-to-category-to-overall macro aggregation, while evaluator code exposes item-to-category-to-overall aggregation.  Historical aggregates are bound only at displayed precision.  Exact values differ while Claude 3 Haiku remains the winner and the two-system order stays fixed.

\paragraph{UCCB.}
The source-explicit 1--5 mean and percentage value differ by the invertible map \(x\mapsto20x\).  They are the same order in different units.  UCCB is therefore a representation/type witness, not headline semantic instability.

\paragraph{BFCL.}
The source exposes official comparison groups and an unweighted language subgroup; an equal ten-group overall is audit-proposed.  The arithmetic closes on 3,399 items, not the 3,981 headline, so the candidate reversal cannot be promoted beyond that subset.

\section{Independent source-table recomputation}
\label{app:independent-recomputation}

These chains read the cited fixed source-table cells rather than downstream discovery outputs.  A second implementation reparses the same displayed cells and reconstructs the orders.  This is independent at source-table level, not raw trajectory replay.

\subsection{MORU}

\begin{table}[ht]
\centering
\caption{MORU exact rational reconstruction; the equal-dimension macro is audit-proposed.}
\label{tab:reviewer-moru-recomputation}
\begin{tabular}{lcc}
\toprule System & Sample-first & Equal 16-dimension macro \\
\midrule
GPT-5.2 & (421/500) & (12593/16000) \\
GPT-5-mini & (102/125) & (3117/4000) \\
Claude Haiku 4.5 & (37/50) & (11413/16000) \\
Gemini 2.5 Flash Lite & (713/1000) & (2157/3200) \\
Grok 4.1 Fast & (711/1000) & (5407/8000) \\
\bottomrule
\end{tabular}
\end{table}

The second reading swaps Gemini and Grok.  GPT-5.2 remains the identified winner and 9/10 pairs are stable.

\Needspace{15\baselineskip}
\subsection{ANIMA}

\begin{table}[ht]
\centering
\caption{ANIMA table-level reconstruction; both columns are source-declared.}
\label{tab:reviewer-anima-recomputation}
\begin{tabular}{lcc}
\toprule System & Historical overall & Dimension-normalised \\
\midrule
Claude Haiku 4.5 & (639/1000) & (8009/13000) \\
Gemini 2.5 Flash Lite & (161/250) & (3971/6500) \\
Grok 4 Fast & (17/25) & (8821/13000) \\
GPT-5-mini & (391/500) & (9681/13000) \\
\bottomrule
\end{tabular}
\end{table}

Claude and Gemini swap; GPT-5-mini remains the winner and 5/6 pairs are stable.  Exactness is relative to the displayed table because raw prompts, grades, and weights are unavailable.

\Needspace{13\baselineskip}
\subsection{BFCL}

\begin{table}[ht]
\centering
\caption{BFCL calculation on the 3,399-item published-category subset.}
\label{tab:reviewer-bfcl-recomputation}
\begin{tabular}{lcc}
\toprule System & Count-weighted micro & Equal ten-group macro \\
\midrule
Claude Haiku 4.5 & (16701/20600) & (5767/7500) \\
GPT-4.1-mini & (225821/283250) & (11639/15000) \\
\bottomrule
\end{tabular}
\end{table}

The order reverses, but the support is 3,399 rather than the 3,981 headline, and the equal-group overall is not source-declared.  This is a support-limited exploratory candidate, not a headline winner flip.

\begin{figure}[ht]
\centering
\begin{tikzpicture}[font=\scriptsize,>=Latex]
  \node[anchor=west,font=\bfseries] at (0,7.2) {MORU: one lower-pair swap; winner fixed};
  \node[anchor=east] (m1a) at (3.2,6.45) {GPT-5.2 (1)};
  \node[anchor=west] (m1b) at (8.0,6.45) {GPT-5.2 (1)};
  \draw[green!55!black,line width=1pt] (m1a.east) -- (m1b.west);
  \node[anchor=east] (m2a) at (3.2,5.85) {Gemini (4)};
  \node[anchor=west] (m2b) at (8.0,5.45) {Gemini (5)};
  \node[anchor=east] (m3a) at (3.2,5.45) {Grok (5)};
  \node[anchor=west] (m3b) at (8.0,5.85) {Grok (4)};
  \draw[orange!75!black,line width=.9pt] (m2a.east) -- (m2b.west);
  \draw[orange!75!black,line width=.9pt] (m3a.east) -- (m3b.west);
  \node[font=\tiny] at (3.2,4.95) {sample-first};
  \node[font=\tiny] at (8.0,4.95) {equal dimensions};

  \node[anchor=west,font=\bfseries] at (0,4.25) {ANIMA: one lower-pair swap; winner fixed};
  \node[anchor=east] (a1a) at (3.2,3.50) {GPT-5-mini (1)};
  \node[anchor=west] (a1b) at (8.0,3.50) {GPT-5-mini (1)};
  \draw[green!55!black,line width=1pt] (a1a.east) -- (a1b.west);
  \node[anchor=east] (a2a) at (3.2,2.90) {Gemini (3)};
  \node[anchor=west] (a2b) at (8.0,2.50) {Gemini (4)};
  \node[anchor=east] (a3a) at (3.2,2.50) {Claude (4)};
  \node[anchor=west] (a3b) at (8.0,2.90) {Claude (3)};
  \draw[orange!75!black,line width=.9pt] (a2a.east) -- (a2b.west);
  \draw[orange!75!black,line width=.9pt] (a3a.east) -- (a3b.west);
  \node[font=\tiny] at (3.2,2.02) {historical overall};
  \node[font=\tiny] at (8.0,2.02) {dimension-normalised};

  \node[anchor=west,font=\bfseries] at (0,1.32) {BFCL: subset winner reverses; support does not close};
  \node[anchor=east] (b1a) at (3.2,.65) {Claude (1)};
  \node[anchor=west] (b1b) at (8.0,.25) {Claude (2)};
  \node[anchor=east] (b2a) at (3.2,.25) {GPT-4.1-mini (2)};
  \node[anchor=west] (b2b) at (8.0,.65) {GPT-4.1-mini (1)};
  \draw[purple!65!black,line width=.9pt] (b1a.east) -- (b1b.west);
  \draw[purple!65!black,line width=.9pt] (b2a.east) -- (b2b.west);
  \node[font=\tiny] at (3.2,-.18) {3,399-subset micro};
  \node[font=\tiny] at (8.0,-.18) {equal ten-group macro};
\end{tikzpicture}
\caption{\textbf{Independent recomputation localizes each order change.} MORU and ANIMA preserve the winner and change one lower pair; BFCL changes the only pair but remains a support-limited exploratory result.}
\label{fig:reviewer-recomputation-slopes}
\end{figure}
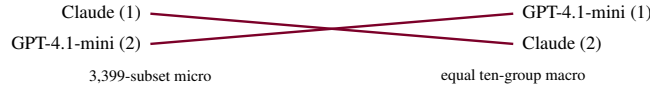

\section{Executable record and deterministic enumeration}
\label{app:executable-contract}

\begin{table}[ht]
\centering
\caption{Minimal field groups for an executable claim-identification record.}
\label{tab:reviewer-input-schema}
\footnotesize
\begin{tabularx}{\linewidth}{@{}p{0.22\linewidth}Y@{}}
\toprule Group & Required content \\
\midrule
Claim and substrate & case/query identifiers; endpoint type, orientation, unit, tie rule; systems and common observation/support keys; scorer, judge, or state binding \\
Candidate ledger & formula or implementation; variation layer; origin; reliability result; A/D/X admission and rationale; same-target/support decision \\
Aggregation and support & denominator and aggregation tree; missingness/DNF rule; weights; unit transform; declared support changes \\
Exposure and freeze & evidence horizon; pre-outcome ledger digest; exposure state; append-only amendments \\
Output or stop & primary/review values, top sets, weak orders, pair relations, witnesses; otherwise stop stage, missing object, no-imputation rationale, and minimum unblock \\
\bottomrule
\end{tabularx}
\end{table}

Origin and admission are distinct.  Admission is executable set construction:
\begin{equation*}
\begin{aligned}
a(f)=\mathrm A&\Rightarrow f\in\Fpri\subseteq\Frev,\\
a(f)=\mathrm D&\Rightarrow f\in\Frev\setminus\Fpri,\\
a(f)=\mathrm X&\Rightarrow f\notin\Frev.
\end{aligned}
\end{equation*}
A source-exposed candidate may remain differently typed; an audit proposal may be executable but review-only; an evaluator implementation failure is routed to a reliability record rather than treated as a new semantic reading.

For each executable record, the analyzer: (1) verifies schema and bound observations; (2) validates endpoint, support, missingness, orientation, units, and ties; (3) evaluates each candidate; (4) projects A into \(\Fpri\) and A+D into \(\Frev\); (5) constructs values, top sets, weak orders, and pair relations; and (6) emits a minimum witness for every non-singleton claim image.  Missing critical fields produce a typed \STOP{} rather than imputation.  The independent verifier recomputes the arithmetic and result profiles through a separate code path.  That establishes mechanical closure only.  Source locators, normalized authorization, same-target reasoning, and A/D/X judgments remain a separate human-auditable provenance layer; the public reproduction package exposes both layers without claiming that deterministic validation proves semantic correctness.

\begin{figure}[ht]
\centering
\resizebox{\linewidth}{!}{%
\begin{tikzpicture}[
  font=\scriptsize,
  >=Latex,
  box/.style={draw,rounded corners=2pt,align=center,text width=2.55cm,minimum height=1.02cm,inner sep=3pt},
  arrow/.style={->,line width=.68pt}
]
\node[box,fill=blue!8] (record) at (0,1.4) {\textbf{Bound record}\\\(D\), query, type, support};
\node[box,fill=blue!8] (validate) at (3.35,1.4) {\textbf{Fail-closed validation}\\orientation, units, ties, missingness};
\node[box,fill=green!9] (project) at (6.70,1.4) {\textbf{A/D/X projection}\\A \(\to\Fpri\); A+D \(\to\Frev\)};
\node[box,fill=purple!8] (enumerate) at (10.05,1.4) {\textbf{Deterministic enumeration}\\values, top sets, orders, pairs};
\node[box,fill=purple!5] (report) at (13.40,1.4) {\textbf{Reader output}\\claim images, witness, backbone};
\draw[arrow] (record) -- (validate);
\draw[arrow] (validate) -- (project);
\draw[arrow] (project) -- (enumerate);
\draw[arrow] (enumerate) -- (report);
\node[box,fill=orange!10,text width=3.35cm,minimum height=.78cm] (stop) at (3.35,-.15) {\STOP{}: first missing object\\+ no-imputation reason + minimum unblock};
\draw[arrow] (validate) -- (stop);
\end{tikzpicture}%
}
\caption{\textbf{Executable contract.} Human judgment fixes type and admission before this state machine.  Thereafter both family projection and identified-set enumeration are deterministic; a missing critical binding exits through a typed stop.}
\label{fig:reviewer-contract-state-machine}
\end{figure}
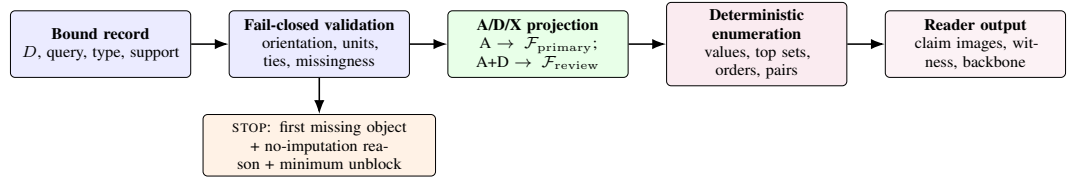

\clearpage
\section{Prospective utility test implied by the census}
\label{app:prospective-utility}

The census is retrospective diagnostic evidence.  It does not estimate how many stops a contract-compliant release would prevent, how much reviewer time it would save, or how often a decision would change.  Those are intervention questions.  A direct next study would register newly released evaluation units and their advertised claims before outcomes, compare a contract-assisted release interface with the legacy interface, and report three separate endpoints: replay completion, time-to-audit, and decision change.  Stop types and minimum-unblock fields provide intervention targets and a baseline taxonomy, not a treatment effect.

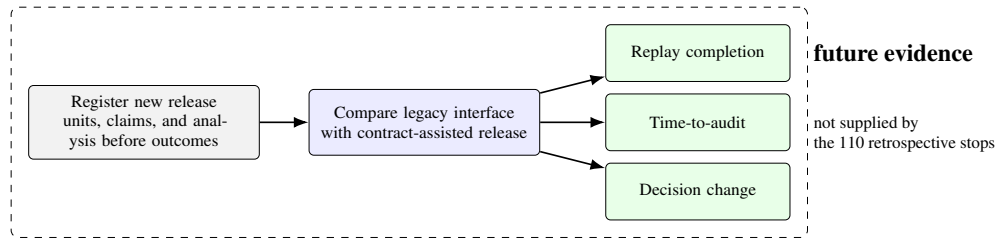
\begin{figure}[ht]
\centering
\resizebox{.94\linewidth}{!}{%
\begin{tikzpicture}[
  font=\scriptsize,
  >=Latex,
  box/.style={draw,rounded corners=2pt,align=center,text width=3.05cm,minimum height=.95cm,inner sep=4pt},
  end/.style={draw,rounded corners=2pt,align=center,text width=2.45cm,minimum height=.82cm,inner sep=3pt},
  arrow/.style={->,line width=.68pt}
]
\node[box,fill=gray!10] (register) at (0,1.25) {Register new release units, claims, and analysis before outcomes};
\node[box,fill=blue!8] (compare) at (4.05,1.25) {Compare legacy interface with contract-assisted release};
\node[end,fill=green!9] (completion) at (8.0,2.25) {Replay completion};
\node[end,fill=green!9] (time) at (8.0,1.25) {Time-to-audit};
\node[end,fill=green!9] (decision) at (8.0,.25) {Decision change};
\draw[arrow] (register) -- (compare);
\draw[arrow] (compare) -- (completion);
\draw[arrow] (compare) -- (time);
\draw[arrow] (compare) -- (decision);
\node[draw,dashed,rounded corners=3pt,fit=(register)(compare)(completion)(decision),inner sep=7pt] {};
\node[anchor=west,font=\bfseries] at (9.55,2.25) {future evidence};
\node[anchor=west,align=left] at (9.55,1.08) {not supplied by\\the 110 retrospective stops};
\end{tikzpicture}%
}
\caption{\textbf{The prospective question is deliberately separate.} The present paper supplies the intervention targets and measurement contract; it does not claim the three future endpoints.}
\label{fig:reviewer-prospective-study}
\end{figure}

\clearpage
\section{Controlled evidence from an orthogonal substrate}
\label{app:legacy-worked-examples}

The following TraceElephant and AgentRx/TELBench controls are not part of the Inspect census and are never pooled with its random, purposive, or exploratory panels.  They provide controlled scientific evidence because they hold predictions or support fixed while varying a declared semantic coordinate.  They show directly what a conventional within-specification uncertainty analysis would miss: an apparently large effect can be localized to a one-sided terminal rule, while non-identified rankings can retain an identified winner or direction.

\subsection{The omitted semantic layer in a controlled diagnostic family}

\begin{figure}[ht]
\centering
\includegraphics[width=.96\linewidth]{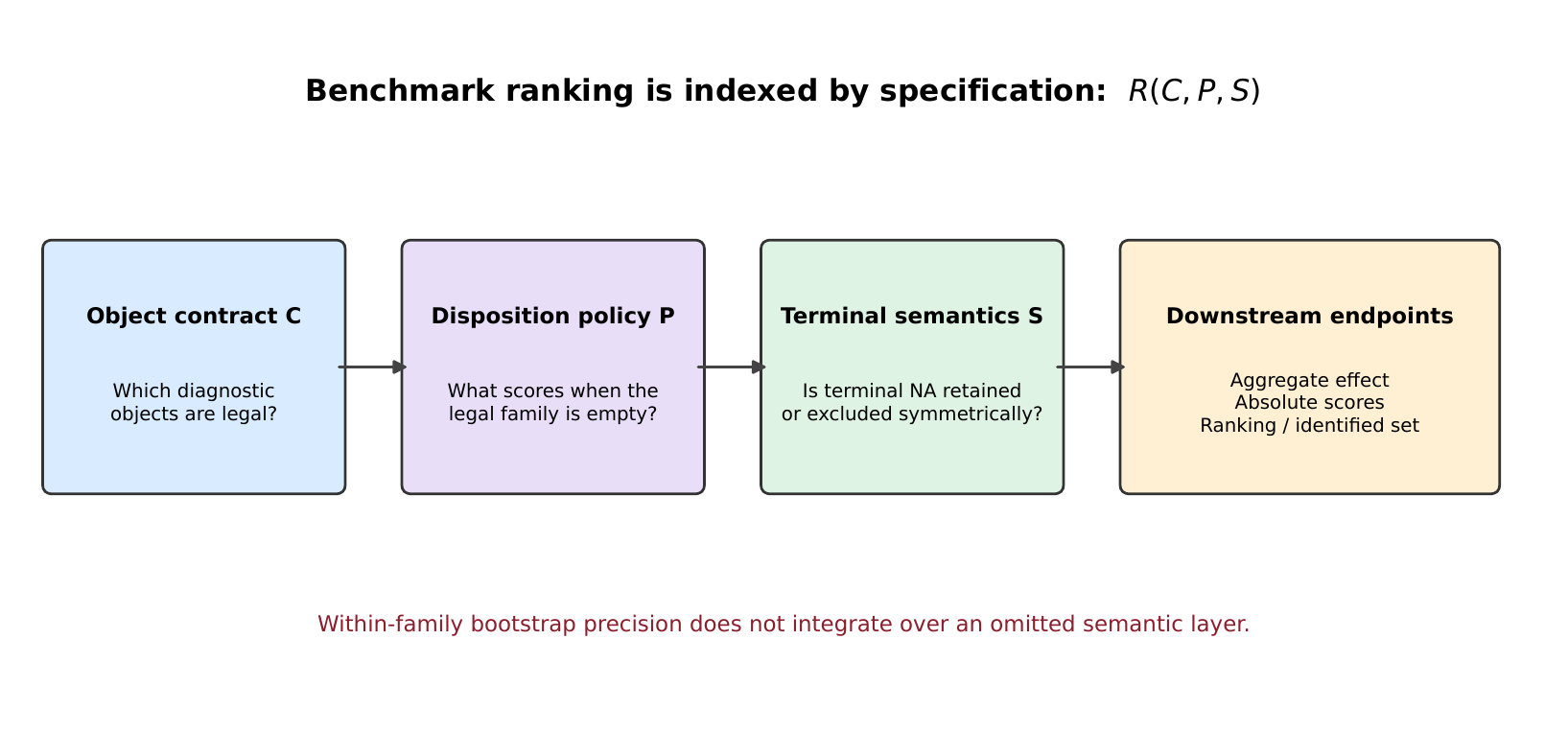}
\caption{\textbf{Controlled diagnostic specifications.} Object contract, disposition policy, and terminal semantics index the downstream endpoint.  Sampling precision inside one box does not integrate over an omitted semantic coordinate.}
\label{fig:reviewer-legacy-layers}
\end{figure}

The common substrate reconstructs 73 AgentRx and 144 TELBench trajectories (217 rows) under seven frozen prediction rules.  Two hundred rows contain multiple candidates and 182 contain multiple canonical carrier structures.  Predictions are never rerun across contracts.  TraceElephant supplies a distinct execution population; its populations and denominators remain separate from the 217-row diagnostic substrate.

For PageRank \(p\) and Earliest Anomaly \(e\), let \(\Delta(s)=Q_p(s)-Q_e(s)\).  A grouped design compares inclusive and adopted-only contracts under a shared disposition/terminal rule and reports their difference in \(\Delta\).  Point scores, rankings, and paired effects are distinct endpoint types.

\subsection{Symmetric terminal controls}

The terminal-neutral retain-both and exclude-both designs were \emph{motivated post hoc, with each control specification frozen before its outcomes were evaluated}.  On the 138-task task-disjoint stratum, every PageRank-minus-Earliest adoption effect is exactly zero under all four backgrounds and three disposition policies.  The original one-sided treatment remains a valid conditional specification but cannot be transported to a terminal-comparable claim.

\begin{table}[ht]
\centering
\caption{Task-disjoint adoption effects and frozen 95\% task-cluster percentile intervals, in percentage points.}
\label{tab:legacy-terminal-effects}
\resizebox{\linewidth}{!}{%
\begin{tabular}{lcccc}
\toprule
Disposition & Original asymmetric & Retain both & Exclude both & \shortstack{Aggregate symmetric\\minus original} \\
\midrule
Empty/empty credit & \(-43.03[-50.82,-35.14]\) & \(0[0,0]\) & \(0[0,0]\) & \(+43.03[+35.14,+50.82]\) \\
Empty no credit & \(-41.30[-48.91,-33.70]\) & \(0[0,0]\) & \(0[0,0]\) & \(+41.30[+33.70,+48.91]\) \\
Conditional nonempty & \(-40.76[-48.61,-32.98]\) & \(0[0,0]\) & \(0[0,0]\) & \(+40.76[+32.98,+48.61]\) \\
\bottomrule
\end{tabular}%
}
\par\vspace{1pt}
{\scriptsize The last column is the endpoint-level contrast between aggregate adoption-effect estimates; each point and interval is generated from the same frozen endpoint binding.  It is not the separately stored mean of per-task direct contrasts.}
\end{table}

The three asymmetric effects differ but are all negative.  Adding symmetric zeros makes strict negativity non-identified while preserving non-positivity.  A four-corner decomposition localizes the apparent effect to deleting terminal candidates on only one contrast side; nonterminal contributions are zero on declared common support.

\paragraph{One-row mechanism.}
On the preserved row, PageRank predicts terminal event \texttt{e0047} and Earliest predicts nonterminal \texttt{e0013}.  The asymmetric treatment deletes PageRank's target only on the adopted-only side, producing row effect \(-1\).  Retain-both scores the two sides \((1,1)\); exclude-both scores \((0,1)\).  Each symmetric treatment therefore gives identical margins across contrast sides without changing predictions.

\begin{figure}[ht]
\centering
\includegraphics[width=\linewidth]{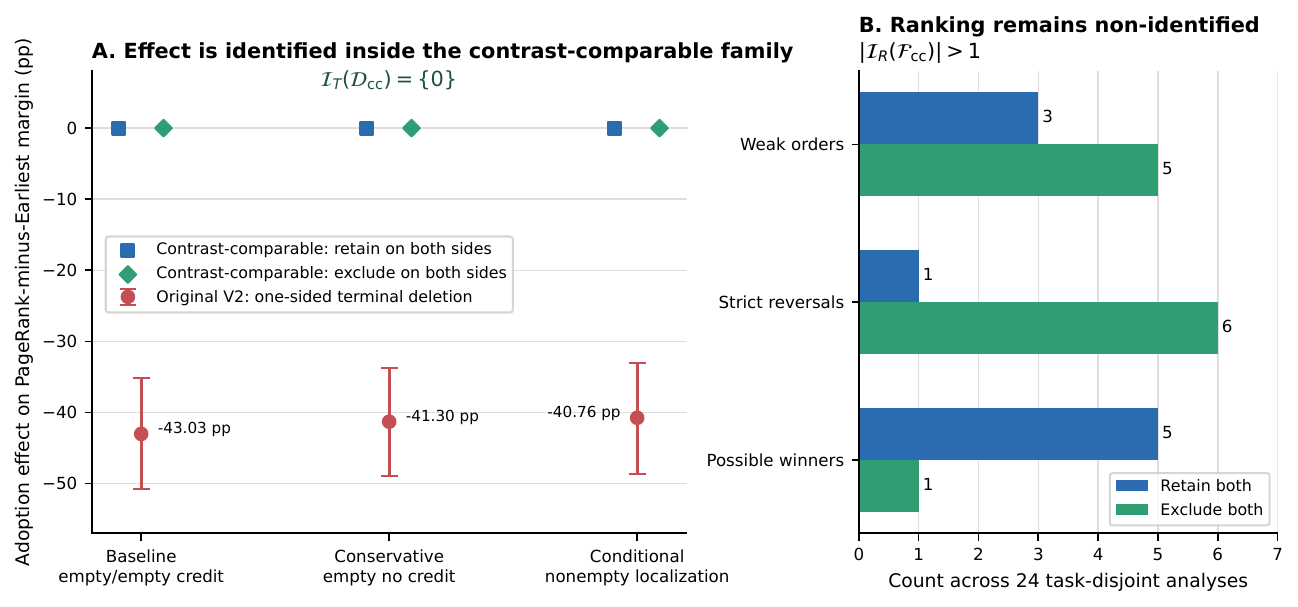}
\caption{\textbf{Controlled result: effect typing and ranking structure separate.} Panel A shows that symmetric terminal handling collapses the contrast-comparable effect to zero.  Panel B shows that complete rankings can remain non-identified while a unique winner or stable relations survive.}
\label{fig:reviewer-legacy-defensive}
\end{figure}

\subsection{Rankings remain non-identified while structure survives}

\begin{table}[ht]
\centering
\caption{Task-disjoint ranking summaries under contrast-comparable terminal treatments.}
\label{tab:legacy-terminal-rankings}
\resizebox{\linewidth}{!}{%
\begin{tabular}{lccccc}
\toprule Treatment & PageRank--Earliest intervals & Weak orders & Reversal pairs & Kendall \(\tau_b\) & Possible winners \\
\midrule
Retain both & \(B/C/N:[43.84,55.43]\pp\) & 3 & 1 & \([0.5534,1]\) & 5 \\
Exclude both & \(B:[-0.72,13.41]; C:[2.54,14.13]; N:[2.54,14.66]\pp\) & 5 & 6 & \([0.4286,1]\) & 1 \\
\bottomrule
\end{tabular}%
}
\end{table}

Exclude-both has five weak orders but Mini is the unique winner in every cell.  In the original eight-contract family, two weak orders occur, 20/21 method pairs are stable, and Mini and Terminal Event Only occupy the top two positions throughout.  A non-singleton complete order therefore coexists with substantial identified substructure.

\subsection{Orthogonal fixed-support empty-family witness}

The second control changes only empty/empty credit while holding all 217 rows, the denominator, terminal treatment, candidate unions, and seven predictions fixed.

\begin{table}[ht]
\centering
\caption{Empty-family semantics on fixed or explicitly narrowed support.}
\label{tab:legacy-empty-family}
\begin{tabularx}{\linewidth}{@{}p{0.27\linewidth}p{0.29\linewidth}Y@{}}
\toprule Policy & Average effect (95\% CI) & Interpretation \\
\midrule
Empty/empty credit & \(-7.72[-12.10,-3.92]\pp\) & Fixed 217-row total-score estimand; cellwise crossing \\
Empty no credit & \(-3.46[-5.99,-1.15]\pp\) & Same rows and denominator; crossing removed \\
Conditional nonempty & \(-0.89[-4.96,+2.87]\pp\) & Narrower support and different endpoint \\
\bottomrule
\end{tabularx}
\end{table}

The fixed-support identified set is \(\{-7.72,-3.46\}\pp\): exact effect non-identified, negative direction identified.  Their paired difference is \(-4.26\pp\) with interval \([-6.68,-2.19]\pp\).  Adding conditional localization produces a broader three-value set but not the same fixed estimand.  The row audit reconstructs \(-67/868=-7.7189\pp\); family collapse accounts for \(-61\) of \(-67\), or 91.0\% in magnitude.

\Needspace{12\baselineskip}
\subsection{Cross-corpus heterogeneity and uncertainty objects}

Under the baseline semantic design, AgentRx gives \(-2.40\pp\) with interval \([-5.82,0]\pp\), TELBench gives \(-10.42\pp\) with \([-17.01,-4.86]\pp\), and the frozen interaction is \(+8.02\pp\) with \([+1.47,+14.95]\pp\).  The corpora are heterogeneous sources, not independent replications.  An independent reconstruction rebuilt 220 raw observable/native-label records, 24,416 row scores, all denominators, margins, weak orders, and 10,000-draw bootstrap arrays; another rebuilt all 868 row-background records.

Finite specification intervals vary declared semantics on fixed observations.  Task-cluster intervals vary resampled tasks under one specification.  Joint sign-pattern frequency is a resampling statistic, not a posterior ordering probability.  A possible-winner set is not a complete-order identified set.  Structural \([0,0]\) intervals hold only for the declared symmetric terminal family.  None of these objects transports to unenumerated semantics without a coverage or invariance assumption.

\end{document}